%% file: main.tex
\documentclass[11pt]{article}

\usepackage{geometry}
\usepackage{setspace}
\usepackage{todonotes}
\usepackage{caption} 
\usepackage{graphicx}
\usepackage{url}
\usepackage{hyperref}
\usepackage{float}
\usepackage{rotating}
\usepackage{mathptmx}
\usepackage[T1]{fontenc}
\usepackage[inline, shortlabels]{enumitem}
\usepackage{cleveref}
\usepackage{multirow}
\usepackage{svg}
\usepackage{pifont}
\usepackage{comment}

\newcommand{\cmark}{\ding{51}}%

\newcommand{\name}{\textbf{CORE metamodel }}

\usepackage{natbib}
\setcitestyle{aysep={}}

\newenvironment{biseabstract}{%
\begin{quote} \bf}
{\end{quote}}

\newenvironment{bisekeywords}{%
\begin{quote} \it \textbf{Keywords:}}
{\end{quote}}

\title{An object-centric core metamodel for IoT-enhanced event logs\footnote{This work has been submitted to the IEEE for possible publication. Copyright may be transferred without notice, after which this version may no longer be accessible.}}

\author{Yannis Bertrand$^{1\ast}$, Christian Imenkamp$^{2}$, Lukas Malburg$^{3}$, Matthias Ehrendorfer$^{4}$, \\
Marco Franceschetti$^{5}$, Joscha Grüger$^{3}$, Francesco Leotta$^{6}$, Jürgen Mangler$^{4}$,  \\ 
Ronny Seiger$^{5}$, Agnes Koschmider$^{2}$, Stefanie Rinderle-Ma$^{4}$, Barbara Weber$^{5}$ and Estefania Serral $^{7}$\\
\\
\normalsize{$^{1}$Department of Business Informatics and Operations Management, Ghent University,}\\
\normalsize{Tweekerkenstraat 2, 9000, Ghent}\\
\normalsize{$^{2}$Group Business Informatics and Process Analytics, University of Bayreuth,}\\ \normalsize{Wittelsbacherring 10, 95444
Bayreuth, Germany}\\
\normalsize{$^{3}$Artificial Intelligence and Intelligent Information Systems, Trier University,}\\ \normalsize{54296 Trier, Germany}\\
\normalsize{$^{4}$Department of Computer Science, Technical University of Munich,}\\ \normalsize{85748 Garching, Germany}\\
\normalsize{$^{5}$Institute of Computer Science, University of St. Gallen,}\\ \normalsize{9000 St. Gallen, Switzerland}\\
\normalsize{$^{6}$Dipartimento di Ingegneria Informatica Automatica e Gestionale, Sapienza Università di Roma,}\\ \normalsize{Via Ariosto, 25, 00185 Rome, Italy}\\
\normalsize{$^{7}$Research Center for Information Systems Engineering, KU Leuven,}\\ \normalsize{Warmoesberg 26, 1000 Brussels, Belgium}\\
\\
\normalsize{$^\ast$Corresponding author: Yannis Bertrand, E-mail: yannis.bertrand@ugent.be}
}

\date{}

\begin{document} 

\baselineskip18pt


\maketitle

\begin{biseabstract}
  Advances in Internet-of-Things (IoT) technologies have prompted the integration of IoT devices with business processes (BPs) in many organizations across various sectors, such as manufacturing, healthcare and smart spaces. The proliferation of IoT devices leads to the generation of large amounts of IoT data providing a window on the physical context of BPs, which facilitates the discovery of new insights about BPs using process mining (PM) techniques. However, to achieve these benefits, IoT data need to be combined with traditional process (event) data, which is challenging due to the very different characteristics of IoT and process data, for instance in terms of granularity levels. Recently, several data models were proposed to integrate IoT data with process data, each focusing on different aspects of data integration based on different assumptions and requirements. This fragmentation hampers data exchange and collaboration in the field of PM, e.g., making it tedious for researchers to share data. In this paper, we present a core model synthesizing the most important features of existing data models. As the core model is based on common requirements, it greatly facilitates data sharing and collaboration in the field. A prototypical Python implementation is used to evaluate the model against various use cases and demonstrate that it satisfies these common requirements.
\end{biseabstract}

\begin{bisekeywords}
Event logs; Internet of Things; Process mining; IoT-enhanced event logs
\end{bisekeywords}



\input{1-intro}
\input{2-running_example}
\input{3-background}
\input{4-methodology}
\input{5-model_presentation}

\input{6-implementation}
\input{7-evaluation}
\input{8-discussion}
\input{9-conclusion}


\setstretch{2.0}
\bibliographystyle{spbasic-bise}
\bibliography{references}


\end{document}

%% file: 1-intro.tex
\section{Introduction}
\label{sec:intro}

The development of Internet-of-Things (IoT) technologies has led to a widespread deployment of IoT devices by organizations to support their business processes (BPs). This trend is observed across many industries, for example, in manufacturing, where sensors and actuators pave the way for process automation \citep{Malburg.2020_BPMAndIoT}; in healthcare, where sensors continuously monitor the vital signs of patients to assist clinical decision-making and enforce best practices in patient handling \citep{fernandez2015process}; in smart spaces, where sensors and actuators support the routines of elderly people \citep{Leotta2015ApplyingPM}.

As discussed by \cite{janiesch2020internet}, the emergence of these so-called IoT-enhanced BPs calls for the development of dedicated business process management (BPM) techniques. These should, for example, coordinate process execution with IoT devices, respond to issues detected by sensors, and provide deeper insights related to the influence of the physical environment on process execution. To achieve these goals, new IoT-enhanced process mining (PM) techniques should be capable of analyzing IoT data together with process events.

One major hurdle preventing the creation of IoT-enhanced PM techniques is the need to integrate IoT data and process event data \citep{bertrand2022assessing,mangler2026internet}. This difficulty is mainly due to the nature of IoT sensor data that is generally at a low granularity level. This in turn means that the data has to be abstracted or filtered to extract relevant information. Moreover, IoT sensor data is noisy and voluminous. This is exacerbated by the fact that IoT devices usually collect data independently of the processes. As a result, relevant IoT sensor data has to be correlated with process data to ensure a meaningful alignment between IoT observations and process events.

To address this integration challenge, several models describing IoT-enhanced event log formats have recently been proposed (e.g., \cite{datastream,bertrand2026nice,franceschetti2023event}). Although they share some similarities, these formats have been designed with different requirements. This is mainly due to 1) each model focusing on specific types of use cases, i.e., different types of processes and IoT devices, which generate different types of data, and 2) each model being optimized for a different data analysis task, with specific input data requirements. For example, DataStream \citep{datastream} assumes a certain degree of process knowledge, typical of manufacturing production processes, and is intended for long-term data storage for auditing. In contrast, CAIRO \citep{franceschetti2023event} is designed for primarily manual and more knowledge-intensive processes. Its main analysis goal is process monitoring in the absence of an information system that tracks processes. In addition to this, they are grounded on different event log standards, i.e., DataStream is an extension of the eXtensible Event Stream (XES) standard \citep{gunther2014xes}, while NICE \citep{bertrand2026nice} follows the object-centric paradigm. This discrepancy between existing IoT-enhanced log formats hinders interoperability and data reuse in research.



To address the limitations of existing formats and foster data exchange and availability for research, a common, unified format is required in the form of a new model, combining the experiences and expertise in this field of research. This common model should capture the main concepts that are shared across the previously proposed formats, and be designed to enable the storage of data from various types of IoT-enhanced BPs. Moreover, it should be possible to easily translate existing logs in previously presented formats to the new common format.

In this paper, we tackle the aforementioned shortcomings by proposing a common metamodel for IoT-enhanced event logs, referred to as \name. This \name specifies the essential concepts needed for IoT-enhanced log storage. The contributions of this paper are manifold and consist of:
\begin{enumerate}[label=\textbf{Contribution~\arabic*}, wide, labelindent=0pt, ref=Contribution~\arabic*]
    \item \label{C1} A set of requirements providing guidance for the development of IoT-enhanced event log formats
    \item \label{C2} A common format for representing IoT-enhanced event logs, i.e., the \name
    \item \label{C3} A prototypical implementation based on the OCEL 2.0 format
    \item \label{C4} Near real-world use cases from several IoT application domains used for evaluation and demonstration
\end{enumerate}
The model builds on previously proposed formats and their requirements for IoT-enhanced event logs (see \ref{C1}). These requirements guide the development process and ensure that the most important concepts of these already available formats are considered to obtain a widely usable standard. By this work, the proposed \name can adequately represent various types of IoT-enhanced event logs and facilitates data storage and exchange (see \ref{C2}). To this end, the model is implemented within the OCEL 2.0 format (see \ref{C3}). This prototypical implementation can be used by other researchers and practitioners to represent IoT-enhanced event logs more easily. In addition, the implementation proposes a transformation procedure to translate already available IoT-enhanced event logs into the new \name format. This translation is demonstrated in our evaluation, in which we use already available IoT-enhanced event logs from several use cases and translate them to the new metamodel format (see \ref{C4}). 

The rest of the paper is structured as follows. First, in Section \ref{sec:background}, we present the relevant background to this paper, covering PM, event data storage and IoT ontologies. Then, Section \ref{sec:method} introduces the methodology we applied to create the \name, following the design science research paradigm. Section \ref{sec:relworks} presents the already proposed formats for IoT-enhanced event logs. Based on these and on the OCEL standard, Section \ref{sec:model} introduces our new metamodel for IoT-enhanced event logs. The implementation that we provide is presented in Section \ref{sec:implementation}, and is used in several use cases to validate our new model, which are described in Section \ref{sec:evaluation}. The model and its implementation are discussed in Section \ref{sec:discussion}, together with their limitations. Finally, Section \ref{sec:conclusion} concludes the paper with a summary of our contributions and ideas for future works.

%% file: 2-running_example.tex
\section{Running example}
\label{sec:example}
\begin{figure}[hbtp]
    \centering
    \includegraphics[width=\linewidth]{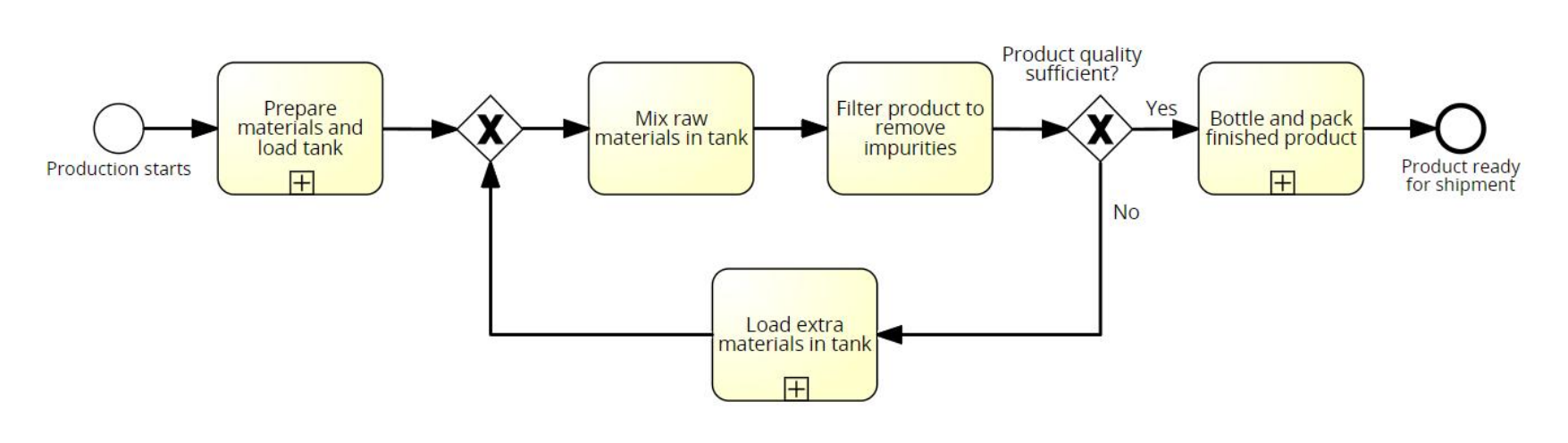}
    \caption{BPMN diagram showing the main steps of the running example process.}
    \label{fig:runexamp}
\end{figure}

To illustrate the need for a specialized format for IoT-enhanced event logs, this section outlines a running real-life manufacturing use case, first introduced in \cite{Bertrand2024Feb}. Figure \ref{fig:runexamp} shows a BPMN diagram of this process.

This use case presents the production process of a company active in the preparation of chemical products. Their production process comprises four main steps: 1) Preparing raw material and loading the tank; 2) Mixing the raw material in the tank; 3) Circulating the product through filters to remove impurities; and 4) Bottling and packing the finished product.

Sensors placed around the process monitor various physical parameters affected by the process, such as the tank pressure and the flow of the product through the filters. The execution of each activity leaves a specific pattern in these IoT data, hence events and activity executions can be detected by analyzing the IoT data. For instance, a sudden peak in product flow may signal a filtering anomaly.

Early detection of flow peaks would enable process engineers to address anomalies early and to prevent the flow from fluctuating too much, which can cause degradation of the quality of the end product. However, this requires analyzing the IoT data from the flow along with process data indicating which activity is currently being executed, a task that is not straightforward. The IoT data are at a low granularity level and need to be abstracted to gain business meaning before they can be combined with process event data~\citep{mangler2026internet}.

%% file: 3-background.tex
\section{Background}
\label{sec:background}
In this section, we present the relevant background to this paper. In Section \ref{sec:PMbackground}, process mining is discussed, first in general terms before focusing on IoT-enhanced process mining and the standards for event log storage. Section~\ref{sec:logsbackground} presents current standards for event logs. Then, Section \ref{sec:ontobackground} presents IoT ontologies, which represent the state-of-the-art in IoT data representation. Section~\ref{sec:relworks} summarizes existing models for IoT-enhanced event logs.

\subsection{Process mining}
\label{sec:PMbackground}
Process mining is a method that leverages event data to analyze and enhance business processes \citep{DataScienceInAction}. It acts as a bridge between the fields of data science and process science \citep{vanderAalst2022,DataScienceInAction}. While data science typically does not prioritize processes and process science focuses mainly on process modeling rather than extracting insights from event data, process mining serves as a crucial tool that seamlessly combines the strengths of both disciplines \citep{DataScienceInAction}.

Process mining encompasses six core techniques, each addressing specific analytical objectives:
\begin{enumerate}
    \item Process Discovery focuses on the automatic creation of process models derived from an event log \citep{van2022foundations}. 
    \item Conformance checking, on the other hand, aims to identify deviations between real-world process events recorded in event logs and a predefined process model \citep{Carmona2018Conformance}. 
    \item Performance analysis addresses challenges such as delays in case completion or missed deadlines. By integrating event data with process models using techniques like token-based replay and alignment methods, it uncovers inefficiencies and bottlenecks \citep{Leemans, vanderAalst2022}. 
    \item Comparative process mining relies on the analysis of multiple event logs, which may differ based on attributes such as location, time period, or resource allocation. This approach allows for the identification of critical similarities and differences between processes~\citep{vanderAalst2022,comparativepm}.
    \item In contrast to these retrospective techniques, predictive process mining combines predictive analytics with process mining methods to forecast future events, outcomes, and remaining time in support of proactive decision-making \citep{Schulte2017Mar,vanderAalst2022,DiFrancescomarino2022Jun}.
    \item Finally, action-oriented process mining bridges the gap between analysis and implementation by translating insights into actionable steps to enhance operational efficiency, e.g. using robotic process automation \citep{Park2022Jul,vanderAalst2022}. 
\end{enumerate}

Applying such techniques to the running use case would enable domain experts to reap insights in the working of the process, e.g., using performance analysis to evaluate the impact of flow deviations on the outcome of the process. However, process mining techniques require a process event log as input and therefore cannot handle raw IoT data. To unleash the power of the IoT, the IoT data first need to be preprocessed and integrated with process event data.

\subsubsection{Process Mining in IoT}
IoT devices generate vast amounts of data reflecting the physical context in which processes operate, offering a rich resource for process mining. Using continuous streams of data from interconnected devices, process mining can move beyond traditional applications to provide context-sensitive, real-time insights into processes, opening new possibilities for advanced techniques \citep{Leotta2015ApplyingPM}. For instance, process mining is increasingly employed in IoT-based systems to detect misbehavior and potential attacks \citep{Hemmer}, as well as to diagnose discrepancies between the designed IoT system and its actual usage. This facilitates anomaly detection and ensures that systems operate as intended \citep{Yamaguchi}.

The time series data produced by IoT devices often includes detailed, low-level information about processes and their executions. This enables the discovery of processes at various levels of abstraction, enhancing event log-driven process discovery by embedding IoT concepts directly into process models \citep{Elkodssi2023Mar}. Moreover, IoT data provides valuable contextual information about the physical environments in which processes occur. Incorporating this context into process mining allows for a more in-depth understanding of the factors influencing process execution, leading to more informed decision-making \citep{Bertrand2024Feb}.

However, integrating IoT data into process mining introduces significant challenges. First, there is a substantial gap between the granular data in sensor logs and the higher-level actions needed for process mining, further complicated by the uncertainty and noise inherent in IoT data, which may lead to ambiguity~\citep{franceschetti2023characterisation}. Second, modeling human behavior is complex due to its unstructured and flexible nature, requiring adaptations such as fuzzy mining or declarative approaches to account for variability and probabilistic patterns. Finally, automatic log segmentation is particularly challenging as sensor logs lack predefined traces and must account for varied routines, temporal constraints, and multi-user interactions, necessitating innovative methods for accurately defining and differentiating human habits \citep{Leotta2015ApplyingPM}.

\subsection{Event log standards for process mining}\label{sec:logsbackground}
\subsubsection{XES}

The current standard event log model is the eXtensible Event Stream (XES) format, an XML-based model that mainly consists of the notions of event, case, and log \citep{gunther2014xes}. It proposes standard attribute types to contextualize the events, e.g. the resource executing an activity, the cost of an activity, etc. A standard activity lifecycle is defined together with XES, based on which the status of an activity can be mapped with events relating to this activity.
XES also allows the definition of new data attribute types through extensions.

\subsubsection{OCEL/OCED}

Recently, the gain in maturity of the PM field has increased the urge to create alternative models to the original XES standard. Multiple propositions that relax some assumptions of XES and allow for more flexibility in event data storage have been presented, e.g., in \citet{popova2015artifact,ghahfarokhi2021ocel,berti2024ocel,fahland2024towards}. In this context, the object-centric event data (OCED) initiative \citep{fahland2024towards} led by the IEEE task force in process mining (TFPM) aims at updating the XES format used by the community with a more flexible standard for event data. The main innovation of OCED is the introduction of the concept of \textit{object}, which generalizes the notion of case by allowing one event to be linked with multiple objects instead of a single case. This removes the necessity to "flatten" the event log by picking one case notion from the several potential case notions that often coexist in real-life processes.

\begin{figure}[hbtp]
    \centering
    \includegraphics[width=.8\linewidth]{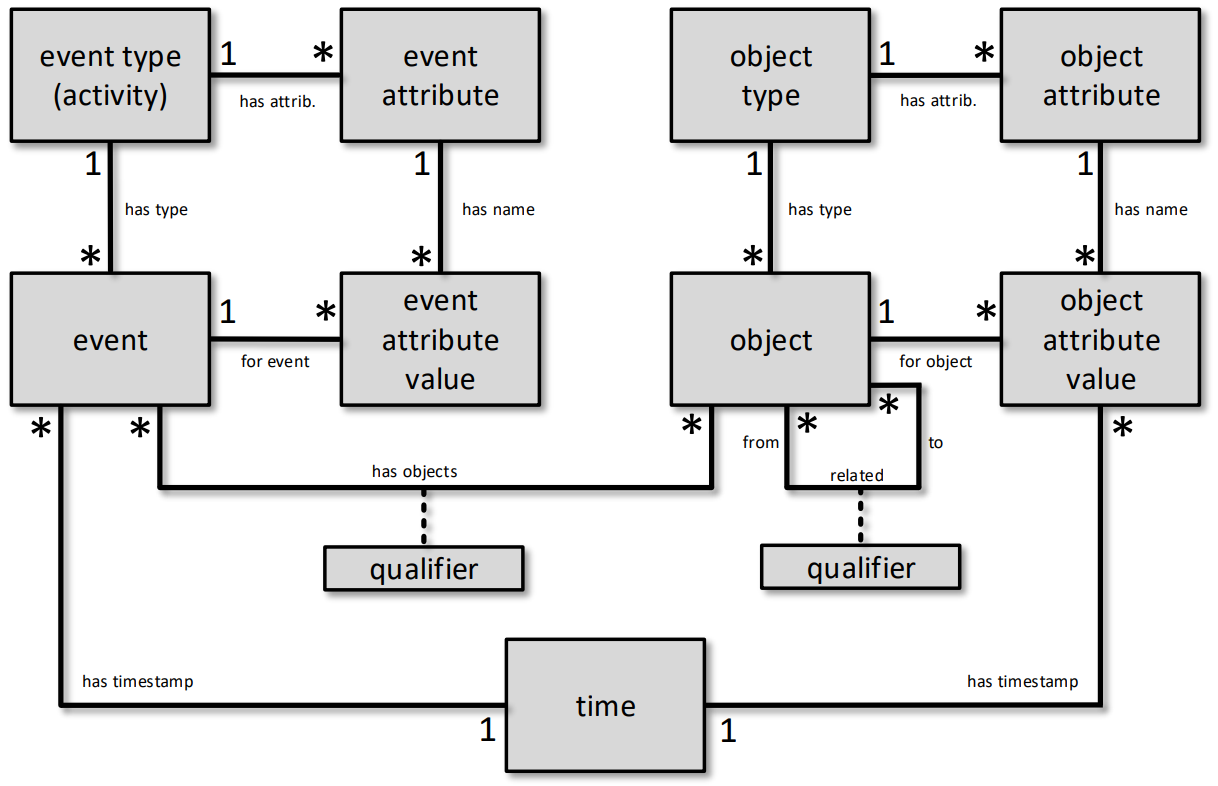}
    \caption{Metamodel of the OCEL 2.0 format (taken from \citep{berti2024ocel}).}
    \label{fig:ocelModel}
\end{figure}

Among the proposed implementations of OCED, the object-centric event log (OCEL) format \citep{ghahfarokhi2021ocel} was designed to be more suitable for storing event logs extracted from relational databases. This model, depicted in Figure \ref{fig:ocelModel}, is currently gaining traction in the PM community as the main contender to replace XES, supported by the availability of event logs and a wide array of software tools enabling the analysis of OCEL event logs. Table \ref{tab:exampleocelLog} shows an example log from the running example in the OCEL format. The first version of OCEL presented in \citep{ghahfarokhi2021ocel} was updated in \citep{berti2024ocel}, which presented OCEL 2.0, solving some of the limitations of the first version of OCEL, such as the static character of object attributes, which limited the capacity of OCEL to store process context.

\begin{table}[hbtp]
    \centering
    \caption{Example OCEL event log fragment.}
    \label{tab:exampleocelLog}
    \begin{tabular}{c|c|cc}
        Event ID & Object ID & Timestamp & Activity label \\
        \hline \hline
        e1 & Batch1, TankA, Matthias & 12:34:56 & Material loading \\
        e2 & Batch1, TankA, Joscha & 12:47:25 & Agitation  \\
        e3 & Batch2, TankC, Matthias & 12:34:56 & Material loading   \\
        e4 & Batch2, TankC, Christian & 12:54:36 & Pump SP adjustment   \\
        e5 & Batch2, TankC, Marco & 13:01:25 & Agitation   \\
        e6 & Batch3, TankD, Marco & 13:17:32 & Filters   \\
        e7 & Batch3, TankD, Lukas & 15:29:25 & Prewet  \\
        e8 & Batch1, TankA, Matthias & 15:37:52 & Filtering  \\
        e9 & Batch1, TankA, Christian & 16:04:16 & Sampling  \\
        e10 & Batch1, TankA, Joscha & 16:08:49 & Material adjustment \\
        \dots & \dots & \dots & \dots
    \end{tabular}
\end{table}

    
\subsection{IoT ontologies}
\label{sec:ontobackground}
Recently, the focus in IoT ontologies has changed from the creation of ontologies that are as complete as possible (e.g. the Semantic Sensor Network (SSN) ontology \footnote{https://www.w3.org/TR/2017/REC-vocab-ssn-20171019/}) to the development of new ontologies that are simpler and more practical (e.g. IoTStream \citep{elsaleh2020iot}). 
Two such ontologies are the Sensor, Observation, Sample and Actuator (SOSA) ontology \citep{janowicz2019sosa} and IoTStream \citep{elsaleh2020iot}.

\subsubsection{SOSA}
SOSA (the Sensor, Observation, Sample, and Actuator ontology) is an otology to formally define sensor networks and their observations \citep{janowicz2019sosa}. At its core, it is composed of three perspectives: (1) Sensors (i.e., physical or virtual) are entities capable of observing properties of features of interest. They derive measurements from other data sources. (2) Observation defines the relationship between different elements (i.e., the sensor, observed property, feature of interest, observation value). They are responsible for tracing and linking the context and origin from the environment. (3) Actuators are entities that define the state change in a feature of interest caused by executing an actuation. 

\subsubsection{IoTStream}
IoTStream is a more specific ontology, inspired by SOSA, that focuses on the treatment of streaming data \citep{elsaleh2020iot}.
The main purpose of IoTStream is to use semantic technologies to bridge interoperability and heterogeneity issues.
These problems arise due to the variety of devices in IoT environments, which also provide different data, making it considerably more difficult to analyze these data and gain advantages. 
Using IoTStream, metadata descriptions can be provided to help search for needed data by representing their properties, such as attribute values and types, finally making data analysis more accessible.

\subsection{Related Work}
\label{sec:relworks}

In this section, we introduce previously developed models for IoT-enhanced event logs. In particular, Section \ref{sec:datastream} introduces the DataStream extension for XES \citep{datastream}; Section \ref{sec:nice} presents the NICE log model \citep{bertrand2026nice}; and Section \ref{sec:cairo} outlines the CAIRO metamodel \citep{franceschetti2023event}.

\subsubsection{DataStream}
\label{sec:datastream}

\begin{figure}[hbtp]
    \centering
    \includegraphics[width=0.7\linewidth]{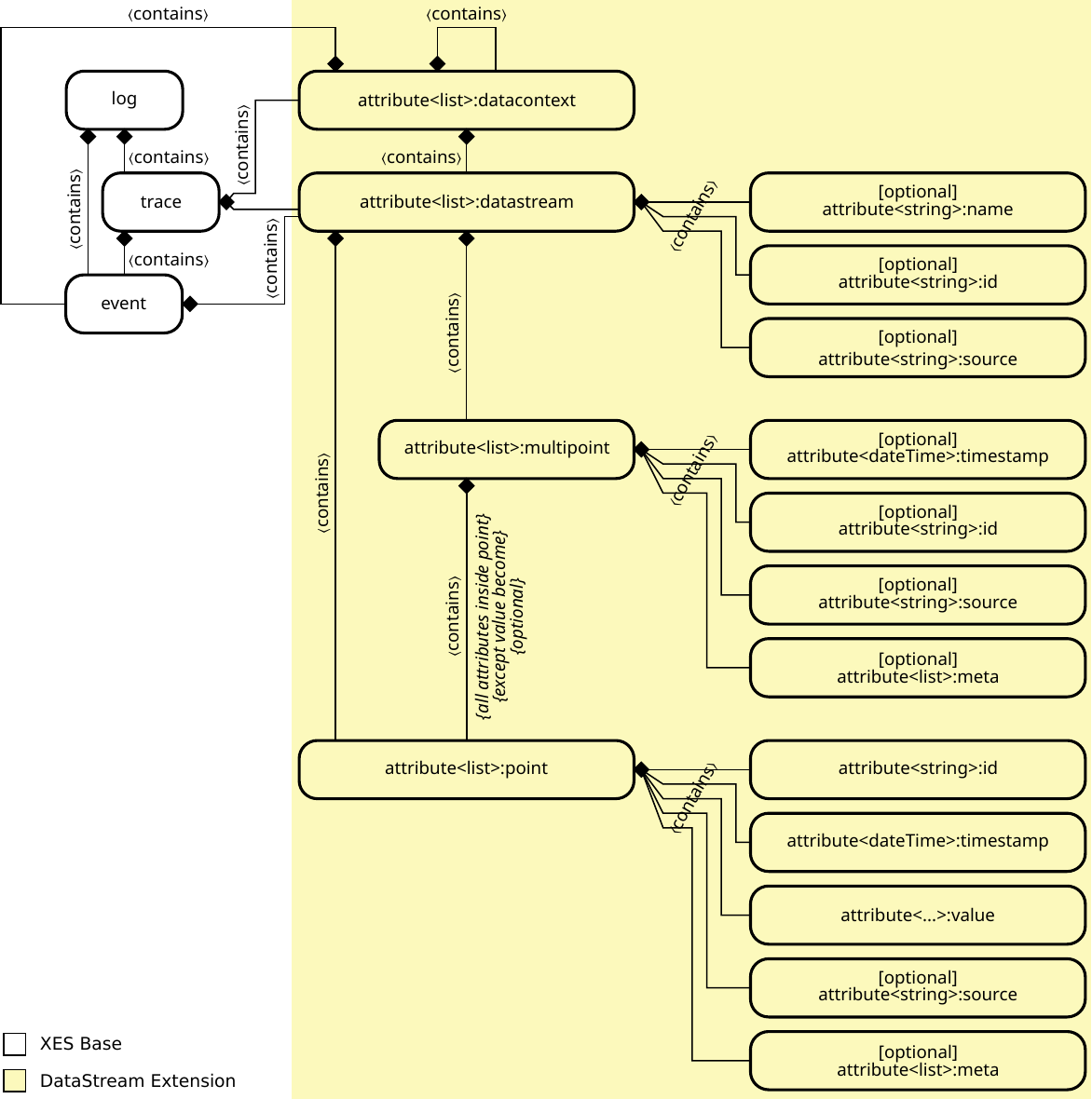}
    \caption{DataStream XES Metamodel~\citep{datastream}}
    \label{fig:datastream}
\end{figure}

The DataStream XES extension~\citep{datastream,Ehrendorfer.2023_IoPTRepository} aims at embedding different kinds of IoT sensor data into eXtensible Event Stream (XES) logs \citep{gunther2014xes}. Therefore, it proposes to integrate additional concepts into the logs (cf. Fig.~\ref{fig:datastream}): \texttt{stream:point} (for representing single measurements), \texttt{stream:multipoint} (for representing multiple measurements with common attributes such as the same id or timestamp), \texttt{stream:datastream} (for grouping multiple points/multipoints together to link them to single events), and \texttt{stream:datacontext} (to provide the possibility to link sensor data to multiple events). By this, it is possible to represent IoT events in addition to process events in trace format. Furthermore, the lifecycle transition \texttt{stream/data} is introduced to allow recording of sensor data independently of the lifecycle events of activities and their control-flow, resulting in sensor data being available at an earlier point in time when monitoring the event stream. This procedure enables the use of the IoT sensor data for checking the conformance of process executions and allows to relate process events to corresponding IoT events and their data streams. The process and IoT events can be stored in a single file, but due to the large amount of sensor data, this is generally not the case. One advantage of DataStream is the possibility to enhance IoT-enriched event logs with semantic annotations so that process or IoT events and their origin can be semantically described in an ontology, for example. In addition, it is possible to store metadata about used data sources in the DataStream format. In contrast, there is no way to handle disambiguity in the log format. In summary, the DataStream XES extension allows for integrating IoT sensor data into XES logs and therefore enables analysis based on the log (ex-post) or based on event streams (at run-time).

\subsubsection{NICE}
\label{sec:nice}

\begin{figure}[hbtp]
    \centering
    \includegraphics[width=.8\linewidth]{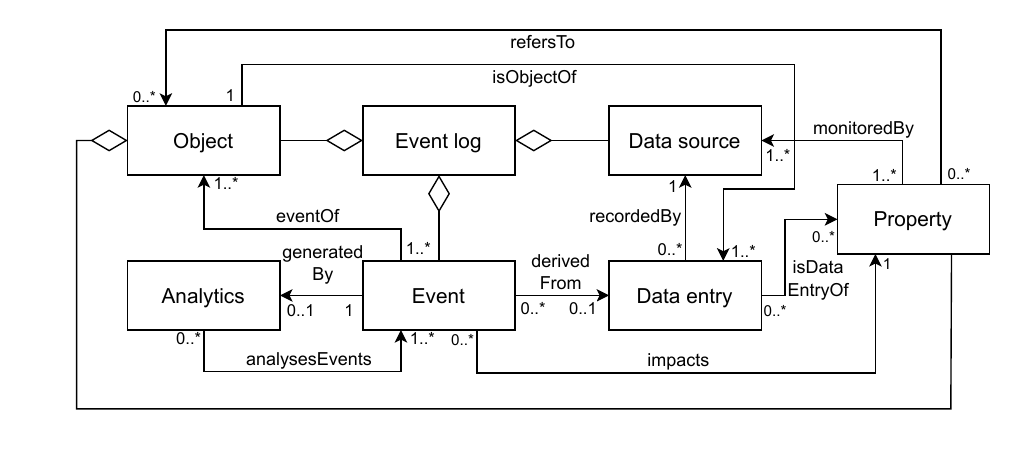}
    \caption{Main concepts of the NICE model \citep{bertrand2026nice}.}
    \label{fig:niceModel}
\end{figure}

NICE \citep{bertrand2026nice} aims at addressing four main challenges in IoT-enhanced event log storage: \textit{i)} the granularity gap between process and IoT data, \textit{ii)} the perspective convolution between control-flow and context information, \textit{iii)} the scope of relevance of the IoT data and \textit{iv)} the dynamicity of IoT data. To do so, NICE proposes a lightweight metamodel comprising seven main concepts, depicted in Figure \ref{fig:niceModel}. At its core, a NICE \textit{Event Log} contains lists of \textit{Data sources}, \textit{Objects} and \textit{Events}. An Event is related to one or several Objects. All Objects can have a collection of \textit{Properties}, which represent context parameters of the process, e.g., the total amount of the order, the temperature in the fridge, the weight and the value of the parcel. Events are derived from one or several \textit{Data entries} or lower-level Events which are logged by a given \textit{Data source}, e.g., an information system (IS) or a sensor. IS can record IS data entries, which typically correspond to rows in a relational database, while sensors make observations, which represent individual measurements of a property of a real-world object.

Object, property, data source and data entry are specialized with subtypes that are digital (i.e., represent the traditional collection of event data by an information system) or physical (i.e., represent the collection of IoT data by IoT devices). This makes it possible to represent the subtleties of data collection in the digital world and in the physical world within a common framework.

Next to this, three types of events are distinguished. First, an \textit{IoT event} represents an instantaneous change in a real-world phenomenon that is monitored by a sensor, or derived from lower-level IoT Events. E.g., the temperature in a fridge decreasing. Second, a \textit{Process event}, which is an instantaneous change of state in the transactional lifecycle of an activity. This corresponds to the traditional notion of event in PM, e.g., a product is taken from a fridge and loaded in a truck. Third, a \textit{Context event} represents an instantaneous change in a property that has an impact on the execution of a specific process instance (i.e., it impacts a Property of an Object) but does not change its control-flow state. 
E.g., the pressure in a tank exceeds a maximum threshold, prompting a human intervention in a chemical production process.

NICE is operationalized with an XML schema, which can be consulted online \footnote{https://github.com/ybertran/NICELogFormat/blob/main/NICELogSchema.xsd}

\subsubsection{CAIRO}
\label{sec:cairo}
\begin{figure}[t]
        \centering
        \includegraphics[width=0.8\textwidth]{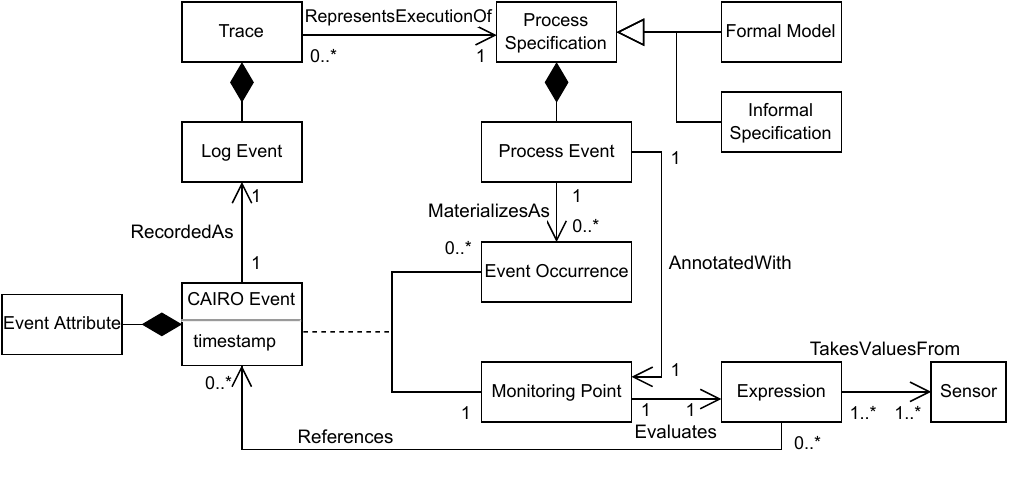}
        \caption{CAIRO metamodel~\citep{franceschetti2023event}}
        \label{fig:cairo}
    \end{figure}

The \emph{CAIRO} metamodel, shown in Figure~\ref{fig:cairo}, has been proposed in~\citep{franceschetti2023event}, motivated by the need for an IoT-driven process monitoring and conformance checking across a wide range of processes and domains. The CAIRO metamodel aims at being agnostic with respect to process and domain characteristics such as level of process structuredness, process modelling paradigm (e.g., imperative or declarative), level of automation of the process activities, and process-awareness, i.e., the support from a Process-aware information system for orchestrating process executions. The foundational assumptions are the observability of process events in terms of IoT events, the knowledge of the monitored process (either formally or informally specified), and a defined mapping of low-level IoT events into high-level process events. At the core of the metamodel is the notion of CAIRO event, i.e., an event that is contextualized, Atomic, Instantaneous, Relevant, and Observable. The contextualized property is represented in the metamodel by the association between the \emph{CAIRO event} class and the \emph{Event attribute} class, which posits that event attributes associate contextual information to an event. The property of being atomic (i.e., events have all-or-nothing semantics) is enforced in the metamodel by cardinality constraints associating each instance of an \emph{Event occurrence} of a CAIRO event with exactly one instance of a \emph{Process event} from a known process specification. The property of being instantaneous binds a CAIRO event to a specific point in time, and is realized by associating a CAIRO event with exactly one \emph{timestamp} attribute. The property of being relevant requires an event to refer to a relevant state change of a process, and is enforced in the metamodel by associating the \emph{CAIRO event} class with the \emph{Monitoring point} class, which models applications detecting activity occurrences based on IoT events. The property of being observable is realized in the CAIRO metamodel with the association between class \emph{Monitoring point} and class \emph{Expression}, representing the aggregation of the readings from instances of the associated class \emph{Sensor}.

\subsubsection{Comparing Existing Models}
Although they share some similarities, the existing models for IoT-enhanced event logs all have some particularities, which distinguish them. The XES DataStream extension enables the storage of all IoT data in the log, together with the impacted events or traces. However, although it is possible to state that a given context is shared by different events within the same trace, there is no straightforward way to represent that a context is shared by events logged in \textit{different} traces. Moreover, the extension only defines two levels of events and does not include any notion of intermediary event. Then, the NICE log format is likely the most flexible model, due to its borrowing the notion of object from the OCEL standard. However, contrarily to DataStream and CAIRO, it does not explicitly enable logging semantic annotations. Finally, CAIRO has similar caveats as DataStream, in that it relies on a more restrictive case notion, but it offers broader support for event contextualization and metadata via event attributes. Hence, a unified model is needed that (i) preserves DataStream's ability to store all IoT data in the log, (ii) inherits NICE's object centric flexibility to span multiple traces, and (iii) retains CAIRO's support for event contextualization and metadata, thereby bridging these existing models.

%% file: 4-methodology.tex
\section{Methodology}
\label{sec:method}

The proposed \name is the artifact resulting from the application of the design science research methodology described in \citep{Hevner.2004_DSRinIS}.
According to this methodology, the development of the artifact represented as the \name follows a sequence of activities.
These activities are based on the Design Science Research Methodology (DSRM) process model by \citep{Peffers.2008_DSRM_ProcessModel}, which consists of six activities:
1) \emph{Identify Problem and Motivate}, 2) \emph{Define Objectives of a Solution}, 3) \emph{Design and Development of an Artifact}, 4) 
\emph{Demonstration}, 5) \emph{Evaluation}, and 6) \emph{Communication}.
This manuscript serves as the medium of \emph{Communication} of the developed artifact. 
The remaining activities are described in the following sections, along with insights on how they have been conducted in the context of the development of the \name.

\subsection{Identify Problem and Motivate}
The development of a common metamodel is motivated by the need to integrate IoT data with process event data in IoT-enhanced business processes--a task complicated by the heterogeneity, noise, and low granularity of IoT sensor data~\citep{bertrand2022assessing}. As shown in Sect.~\ref{sec:background}, current models differ significantly in their assumptions, application domains, and design objectives, resulting in fragmentation that impedes interoperability and data reuse. A common metamodel aims to bridge these gaps by capturing core concepts shared across existing models, facilitating translation and standardization of logs across diverse use cases. This promotes broader adoption by fostering the exchange of IoT-enhanced process data.

\subsection{Define Objectives of a Solution}
\label{subsec:define_objectives_of_a_solution}
To develop a suitable research artifact, concrete objectives for the solution have to be defined.
Here, the objective is to provide a common metamodel for the representation of IoT sensor data and business process data in a unified way.
For this purpose, requirements have been derived from multiple research groups that work in the field at the intersection of IoT and BPM. These research groups have already developed their own models to represent IoT-enhanced event logs based on specific requirements.
An initial set of requirements has been collected from the already available models by analyzing their respective requirements, identifying commonalities, and discussing priorities for a common metamodel. Afterward, this set of requirements has been refined through several discussion rounds, finally resulting in seven requirements that are the basis for the development of the \name.
\begin{enumerate}[label=\textbf{R\arabic*}, wide, labelindent=0pt, ref=R\arabic*]
	\item \label{R1} \textbf{-- Different Granularity Levels:} Different granularity levels must be representable, including a) IoT events (e.g., low-level sensor data, higher level IoT events~\citep{seiger2023data}) and b) process events (e.g., state changes in workflow executions).

    \item \label{R2} \textbf{-- Context Independent of Control-flow:} Context data, like IoT data, is not always directly tied to a specific event of the process but is crucial for understanding process instances. Therefore, IoT data should be loggable independently of process events to ensure comprehensive context representation.

    \item \label{R3} \textbf{-- Traceability:} Traceability across different granularity levels is essential. Therefore, it should be possible a) to establish relationships between low-level and high-level events, such as linking IoT events to process events. Furthermore, since event abstraction can be part of functions or algorithms, b) it should be possible to annotate an event with the abstraction algorithm or expression used, indicating that it was generated through this process \citep{GonzalezLopezdeMurillas2019}.


    \item \label{R4} \textbf{-- Semantic Annotations:} In semantics-aware process analytics~\citep{rebmann2022enabling}, prior knowledge can be used, if available. Therefore, the metamodel should be able to represent a) procedural knowledge, such as information about the process model (formal or informal, e.g., in natural language), b) the relationship between sensors and the process within the infrastructure, and c) prior knowledge from other data sources, such as databases or taxonomies. 

    \item \label{R5} \textbf{-- Flexible Case Notion:} For some use-cases, a single case notion is not sufficient. Therefore, a more flexible case notion is desired, which allows to not only link events to one specific case but to (potentially multiple) entities being involved in its execution. This enables analysis from different points of view (i.e., based on different corresponding entities such as machines, locations, etc. \citep{GonzalezLopezdeMurillas2019}.

    \item \label{R6} \textbf{-- Different Data Sources:} Data in an IoT environment can originate from different sources, which are very heterogeneous. It should be possible to integrate and represent various data sources, including a) information systems (IS), such as ERP systems or databases containing process metadata, b) sensors, which provide real-time data like temperature, humidity, or motion, and c) actuators, which record actions such as opening a valve, starting a motor, or adjusting a device’s settings \citep{GonzalezLopezdeMurillas2019}. Moreover, it should be possible to represent these sources in a unified manner, despite the high level of heterogeneity.

    \item \label{R7} \textbf{-- Represent Metadata:} 
    The log should also include metadata that describe the broader context or configuration in which events are recorded. Unlike semantic annotations, metadata applies to the log or logging setup as a whole rather than individual events. Typical examples include: a) Information about the data acquisition environment (e.g., machine configurations, sensor firmware versions, log generation settings); b) Timestamps or durations relating to log generation periods (e.g., start/end of logging, time zone offsets); and c) Global properties relevant to interpretation (e.g., logging strategy). This metadata is essential for ensuring correct interpretation, comparison, and reuse of event logs across systems and settings.
    

\paragraph{Requirements Fulfillment}
To highlight the need for a common metamodel, Table \ref{tab:reqComparison}, provides an overview of the fulfillment of the above requirements by the models presented in Section \ref{sec:relworks}. 
It is clear from the table that each existing model has varying degrees of fulfillment of the requirements, which align with the specific goals driving their development. For instance, NICE is designed to be very flexible, which translates into fulfilling \ref{R2} and \ref{R5}, while CAIRO focuses on knowledge-intensive processes, as evidenced by its fulfillment of \ref{R4} and \ref{R7} via event attribute, monitoring point, and process specification.

\begin{table}[hbtp]
    \begin{center}
        \caption{Comparison of previously proposed models with respect to the requirements.}

        \begin{tabular}{lccc}
        \multirow{2}{*}{Requirements} & \multicolumn{3}{c}{Preexisting model} \\
         & DataStream & NICE & CAIRO \\ \hline
         R1 – Different Granularity Levels & \cmark & \cmark & \cmark \\
         R2 – Context Independent & \multirow{2}{*}{} & \multirow{2}{*}{\cmark} & \multirow{2}{*}{} \\
          from Control-flow & & & \\
         R3 – Traceability & \cmark & \cmark & \cmark \\
         R4 – Semantic Annotations & \cmark & & \cmark \\
         R5 – Flexible Case Notion & & \cmark & \\
         R6 – Different Data Sources & \cmark & \cmark & \cmark \\
         R7 – Represent Metadata & & & \cmark \\
        \end{tabular}
    \label{tab:reqComparison}
    \end{center}
\end{table}

\end{enumerate}

\subsection{Design and Development of an Artifact}
The development of the \name required a structured and scientifically grounded alignment process among international experts from six specialized working groups. To achieve this, a participatory and iterative development approach has been adopted, drawing on the principles of Participatory Design \citep{bodker2022participatory} and iterative prototyping \citep{camburn2017design}. This methodology ensured the active participation of all stakeholders while allowing continuous validation and refinement of the model.

The working groups have been represented by 1 to 2 researchers in a series of workshops, where initial drafts of the metamodel were collaboratively developed and critically discussed. Key activities during these sessions included evaluating alternative modeling approaches, assessing alignment with previously defined requirements, and validating the model against specific use cases. The output of each workshop consisted of one new draft of the metamodel, which has been subsequently reviewed within the respective local research groups. During this phase, the drafts were tested and refined against the groups' individual use cases, allowing for comprehensive feedback collection and further contextual validation.

The aggregated input from these decentralized discussions was brought back to subsequent workshops, where it informed the iterative refinement of the model. This sequence of activities--a workshop followed by local group discussions and subsequent model adjustments--has been repeated in eight iterations, each building on the feedback and insights gathered in the previous rounds. This iterative approach not only facilitated the integration of diverse perspectives, but also ensured the evolving metamodel's practical applicability across different domains and scenarios.

\subsection{Demonstration}
In order to demonstrate the appropriateness of the developed \name and as a further contribution of our paper, the proposed \name has been implemented as an OCEL 2.0 extension. The implementation serves as a proof-of-concept and allows other researchers to use the \name for their own research.

\subsection{Evaluation}
Based on the proof-of-concept implementation, an evaluation has been conducted to assess the expressiveness of the \name. The primary objective of the evaluation aimed to verify its ability to comprehensively capture the relevant concepts of different preexisting metamodels and fulfill the common requirements. To this end, already available IoT-enhanced event logs represented in one of the different source metamodels discussed in Sect.~\ref{sec:background} have been translated into the proposed \name. The evaluation focused on the identification of potential information loss during the translation process, as any loss would signify a gap in the expressiveness of the \name. As the metamodel is designed to represent a wide range of IoT-enhanced event logs, the evaluation encompassed concrete event logs from different use cases. For this purpose, IoT-enhanced event logs have been used, which have been provided by the working groups that proposed the metamodels discussed in Sect.~\ref{sec:background} and that build at the same time the basis for the development of the \name. The evaluation is presented in detail in Sect.~\ref{sec:evaluation}.

%% file: 5-model_presentation.tex
\section{Model presentation}
\label{sec:model}

In this section, we present the \name. This metamodel is the result of the workshops with the working groups and integrates the core concepts common to the metamodels and use cases of each working group, fulfilling the requirements presented in Section \ref{subsec:define_objectives_of_a_solution}.


\subsection{Main Concepts Derivation from Existing Models}
\label{subsec:link_to_other_models}

To create the common \name, we started by analyzing existing models to 1) identify similarities and 2) distinguish features that addressed the common requirements (necessary features for the common model; in some cases decide which solution to keep) from features that are specific to the models' use cases or features that violate the requirements (e.g., single case notion in DataStream).

\begin{table}[!ht]
    \centering
    \caption{Overview of the mapping of concepts from pre-existing models. Note that attributes and child concepts which were modeled similarly as their parent concept were omitted for simplicity.}
    \begin{tabular}{lll}
        \textbf{Pre-existing concept} & \textbf{IoT-enhanced log models} & \textbf{Concept in our model} \\ \hline
        \hline
        Log & DataStream, NICE & / \\ \hline
        Trace & DataStream, CAIRO & / \\ \hline
        Event & DataStream, NICE & \multirow{2}{*}{Process event} \\
        Event Occurrence & CAIRO & \\ \hline
        DataContext & DataStream & / \\
        DataStream & DataStream & Link object (top-down) \\ \hline
        MultiPoint & DataStream & / \\ \hline
        Point & DataStream & \multirow{3}{*}{IoT event} \\
        Observation & NICE & \\
        CAIRO Event & CAIRO & \\ \hline
        \hline
        Object & NICE & Business object \\ \hline
        Data source & NICE & Data source object \\ \hline
        Sensor & NICE, CAIRO & Sensor object \\ \hline
        Information system (IS) & NICE & IS object \\ \hline
        Analytics & NICE & Link object (bottom-up) \\ \hline
        Data entry & NICE & / \\ \hline
        Property & NICE & / \\ \hline
        \hline
        Process specification & CAIRO & / \\ \hline
        Process event & CAIRO & / \\ \hline
        Monitoring point & CAIRO & / \\ \hline
        Expression & CAIRO & / \\ \hline
    \end{tabular}
    \label{tab:conceptsmap}
\end{table}

Then, in order to  1) reuse tooling and 2) ensure visibility, we integrated these concepts within the OCEL 2.0 metamodel. The main reason for linking our model with OCEL 2.0 is its growing popularity in the process mining community, among both researchers and practitioners, which we think will be crucial to enhance the visibility and uptake  of our model. Moreover, this allows us to reuse the broad software tooling readily available to manipulate and analyze object-centric event data. Concretely, this means that we translated the common core concepts in either \textit{events} and \textit{objects}, the two main concepts in the OCEL 2.0 metamodel. 

We therefore refined event and object into explicit subtypes to precisely define the events and objects used in IoT-enhanced event logging. 

Finally, we derived relationships between the different subtypes of objects and events that we defined. To do so, we started from the OCEL 2.0 metamodel, which defines the two following relationships:
\begin{itemize}
    \item the event-object relationship, which has a many-to-many cardinality;
    \item the object-object relationship, which has a many-to-many cardinality.
\end{itemize}

Note that there is no direct event-event relationship; events can only be related indirectly, via a common object. To obtain the relationships in the \name, we adapted the relationships in existing models to respect theses constraints from the OCEL 2.0 metamodel.

In the remainder of this section, we first present a high-level view of our metamodel, before diving deeper into how we modeled the concepts retained from the original IoT-enhanced log models. Then, we give a concrete example of use of the \name from the running example. 

\subsection{Metamodel Presentation}
\label{subsec:metamodel_presentation}
\begin{figure}[hbtp]
    \centering
    \includegraphics[width=\linewidth]{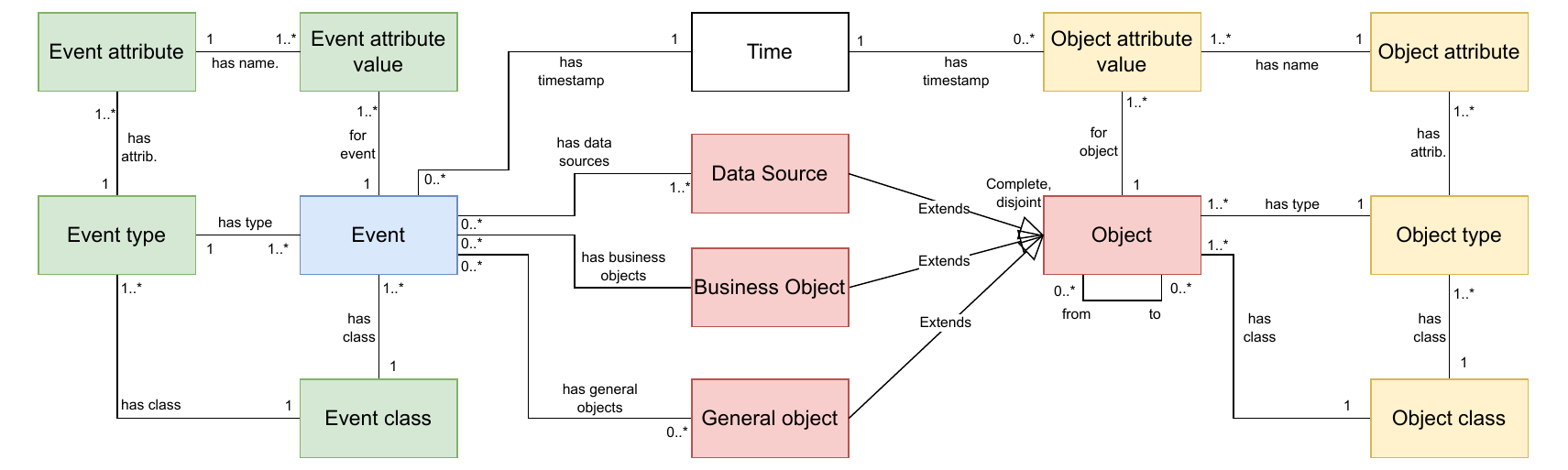}
    \caption{High-level view of the metamodel.}
    \label{fig:highLevelMetamodel}
\end{figure}

The metamodel, which is represented as a UML class diagram in Figure \ref{fig:highLevelMetamodel}, consists of two main concepts, rooted in the object-centric paradigm:
\begin{itemize}
    \item \emph{Events} (in blue), which are defined as happenings that are relevant to a process or its context and are linked to related information (in green);
    \item \emph{Objects} (in red), which are defined as digital artifacts or real-world objects (or a combination of both) that play a role in the execution of a process and are linked to related information (in yellow).
\end{itemize}


Note that, in addition to objects and events, we also include \emph{Time} for consistency with OCEL 2.0.


As can be seen in the UML class diagram in Figure~\ref{fig:highLevelMetamodel}, we divide objects into three main classes: 
\begin{itemize}
    \item Data sources: system that collects data, such as IS or IoT devices that collect events;
    \item Business objects: object of interest for the process, which are impacted by events. They can correspond to process models, cases, or they can define the scope of a context parameter;
    \item General objects: other types of objects, which do not fall into one of the two previously introduced classes, are explicitly mentioned for emphasis. Examples include subprocesses or activity objects \citep{benzin2024inexa}.
\end{itemize}

Each event needs exactly one data source and one or more business objects, and can (but does not have to) be linked to objects of other classes.

More specifically, we distinguish two classes of events: \emph{IoT events} and \emph{process events}.

\subsubsection{Event Type Definitions}

\label{subsec:event_type_definition}
In process mining, an event is commonly defined as the execution of (part of) an activity \cite{gunther2014xes}. However, an event has a very different meaning in the IoT literature, where it is often seen as an observation made by a sensor \cite{janowicz2019sosa, Bertrand2024Feb}. To resolve this discrepancy, our model introduces two types of events, \textit{IoT events} and \textit{process events}, and gives them explicit definitions. 

An IoT event represents a change in a physical phenomenon, which is typically derived from one observation of one IoT device such as a sensor, e.g., a measurement made by a temperature sensor. To account for complex physical phenomena, which cannot directly be tracked by an IoT device, but require to process the observations made by one IoT device or to aggregate data collected by multiple IoT devices, we also make it possible to derive a higher-level IoT event from one or more lower-level IoT events \cite{seiger2023data}, which can be characterized by a label indicating the semantics of the event. An example of such an IoT event is the  detection of a peak in time series sensor data, which requires processing consecutive observations of one sensor using an algorithm.

Next to this, a process event is defined as a change in the lifecycle of a process activity, which can be directly logged by a (process-aware) information system or can be derived from several IoT events. Examples of process events include the registration of a new order, the notification to a customer that a claim was handled, and the arrival of a parcel at a distribution center. In the running example (cf.~\ref{sec:example}), process events relate to activities such as loading the tank, filtering the product, and packing the product.

\begin{table}[htbp]
    \centering
    \caption{Summary of the characteristics of both types of events defined in the model.}
    \begin{tabular}{c|c|c}
       \textbf{Characteristic} & \textbf{IoT event} & \textbf{Process event} \\
       \hline
       Physical vs Digital & Physical & Physical or digital \\
       Granularity level & Low & High \\
       Activity relationship & Relates to 0-n activities & Relates to 1 activity \\
       Result & Observation & Activity label \\
       Aggregation & Consists of 1-n observations & Consists of 1-n IoT events \\
    \end{tabular}
    \label{tab:eventCharacteristics}
\end{table}

\subsubsection{Objects Taxonomy}
\label{subsec:objects_taxonomy}
\begin{figure}
    \centering
    \includegraphics[width=0.9\linewidth]{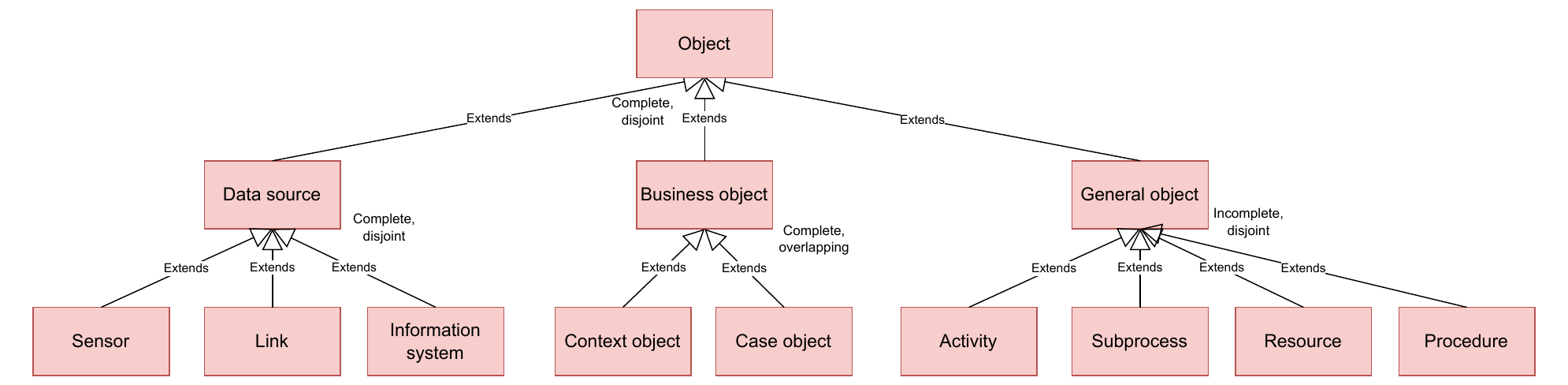}
    \caption{Taxonomy of objects defined in our core metamodel.}
    \label{fig:objtaxonomy}
\end{figure}

In this section, we further refine the main object classes described above, defining more specific object types related to different event types, and making the link between them. 
Figure \ref{fig:objtaxonomy} depicts our taxonomy of object types.


\paragraph{Data sources}
Three mutually exclusive types of data sources are defined: \textit{Sensors}, \textit{Information Systems (IS)}, and \textit{Links}. 

First of all, \emph{Sensors} are IoT devices present in the physical environment of business processes. They collect data in the form of IoT events, which give information about the context of a process. Sensors can therefore be linked to 0 to many IoT events (see Figure \ref{fig:IoTProcEventsOCEL}). Examples include flow sensors monitoring the flow of the product, temperature sensors that track the temperature in the refrigerated area where raw materials are stored. 

Then, \emph{IS} are systems that support the execution of BPs, typically process-aware IS, ERP, MES, etc. An IS records process events and can be linked to 0 to many process events.

Finally, \emph{Links} are optional objects that can be used to represent a relation or a dependency between lower-level events and higher-level events. Such a relationship can be of two types: 1) a link can specify that a process event was derived from one or several IoT events, or an IoT event was derived from one or several lower-level IoT events (bottom-up link; in the running example a bottom-up link could represent the event abstraction method or system used to derive a peak in the flow of the product from consecutive observations of the flow, e.g., complex event processing); 2) a link can represent that an IoT event was prompted by a process event, e.g., because an activity execution requires fetching the value of a Sensor (top-down link; in the running example, a top-down link could fetch extreme values of the flow during the assessment of the quality of the product and compare them to threshold values to determine whether an adjustment is necessary). 


\paragraph{Business objects}
In our metamodel, we distinguish between two potentially overlapping types of business objects: \textit{Case objects} and \textit{Context objects}.

First, the Case objects represent traditional objects as defined by the OCED \citep{fahland2024towards} and OCEL standards \citep{berti2024ocel}. Such objects are entities that are involved in the execution of activities. As such, they represent cases in subprocesses that provide different views on long end-to-end processes. Examples include orders, product bottles, batches, etc.

Next, we define Context objects as objects that have attributes of interest to the process. For our purpose, this will most of the time represent physical objects (or Features of Interest in SOSA) to which IoT devices are attached to track specific parameters. For example, a tank that is equipped with sensors to measure the flow of the product is a context object; another example is the clean room where the process is executed and where temperature, humidity, or the presence of a person is monitored by IoT devices.

Note that these object types can potentially overlap: a tank can also be a case object if it is involved in some specific process activities, or a batch can have some attributes of interest such as product flow tracked by IoT devices.


\paragraph{General object}
Finally, next to the Data sources and Business objects, we represent other types of objects as General objects, for emphasis. These objects may be present in the log and have a specific meaning, but they are not required and do not play a role in the integration of IoT and process event data. Examples of general objects include (but are not limited to):

\begin{itemize}
    \item Activity objects, which are used to group process events that refer to the same activity execution (typically, start and complete events);
    \item Subprocess objects, which can group process events that represent a coherent part of the business logic of the process, but do not have a data entity (case object) in common (e.g., the series of steps necessary to load a raw material in the tank in the running example);
    \item Resource objects, which represent the worker (or machine) that executes one or several activities;
    \item Objects related to the collection of IoT data, i.e., from the SOSA ontology, such as the procedure used to perform data collection in the tank;
\end{itemize}

\subsection{IoT and Process Events Representation}

\begin{figure}[H]
    \centering
    \includegraphics[width=\linewidth]{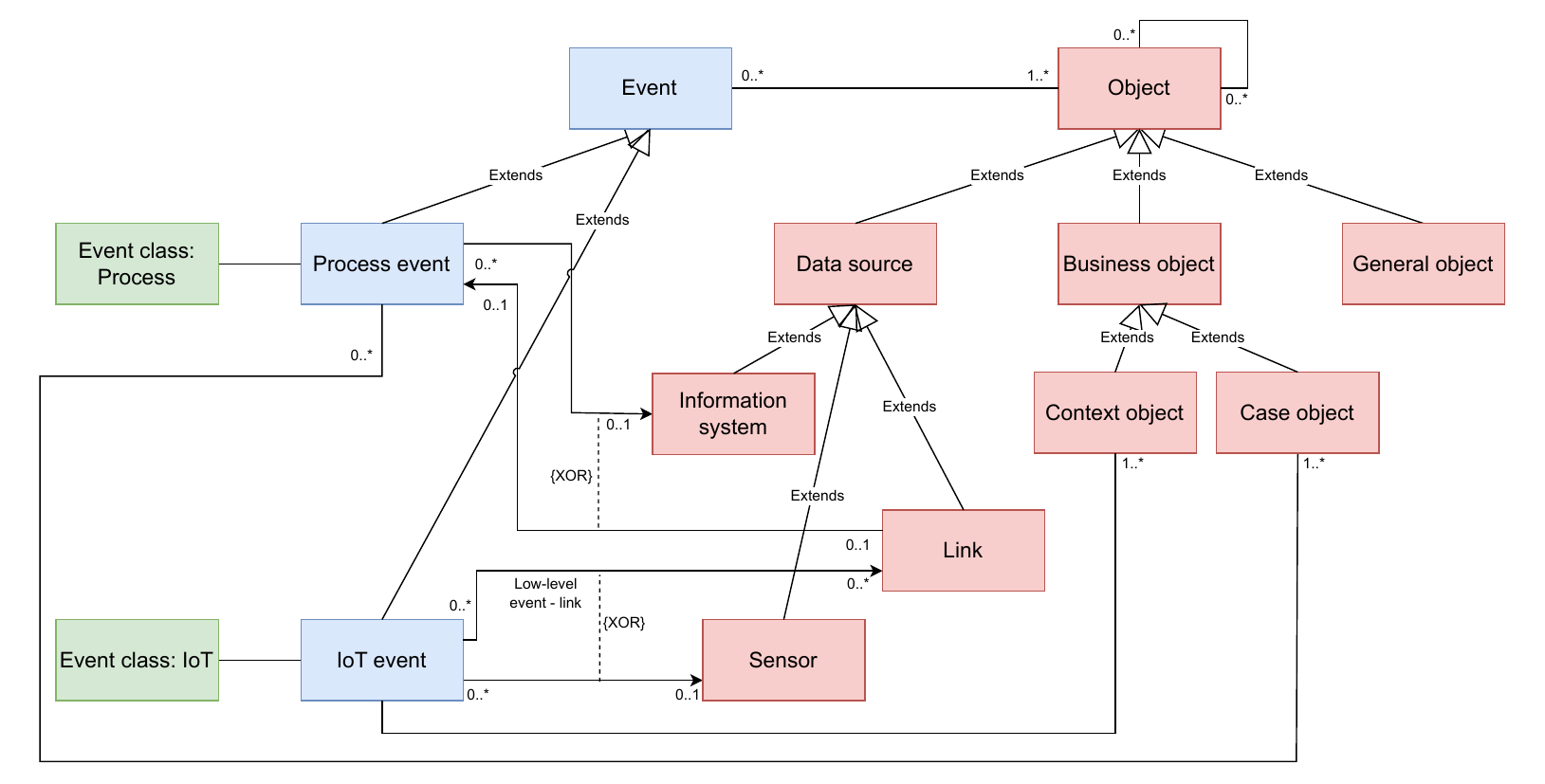}
    \caption{IoT and process events modeling in OCEL.}
    \label{fig:IoTProcEventsOCEL}
\end{figure}

IoT and process events can be modeled as represented in Figure \ref{fig:IoTProcEventsOCEL}. Observations of an attribute of a Context object are made by a Sensor, which logs the observations as IoT events, with the value of the physical parameter represented as an attribute of the IoT event.


Subsequently, multiple IoT events can be used by a Link object to derive a higher-level event (either IoT or process event).


Accordingly, Process events can have two Data sources: an IS or a Link. In the first case, the event is logged by the IS at runtime, along with possible event attributes. In the second case, the process event is derived from one or multiple IoT events, via a Link object, which records the method or rule used to derive the process event. In both cases, the process event is related to one or more Case objects, indicating the process instance it belongs to.


\subsection{Example}
\begin{figure}
    \centering
    \includegraphics[width=.8\textwidth]{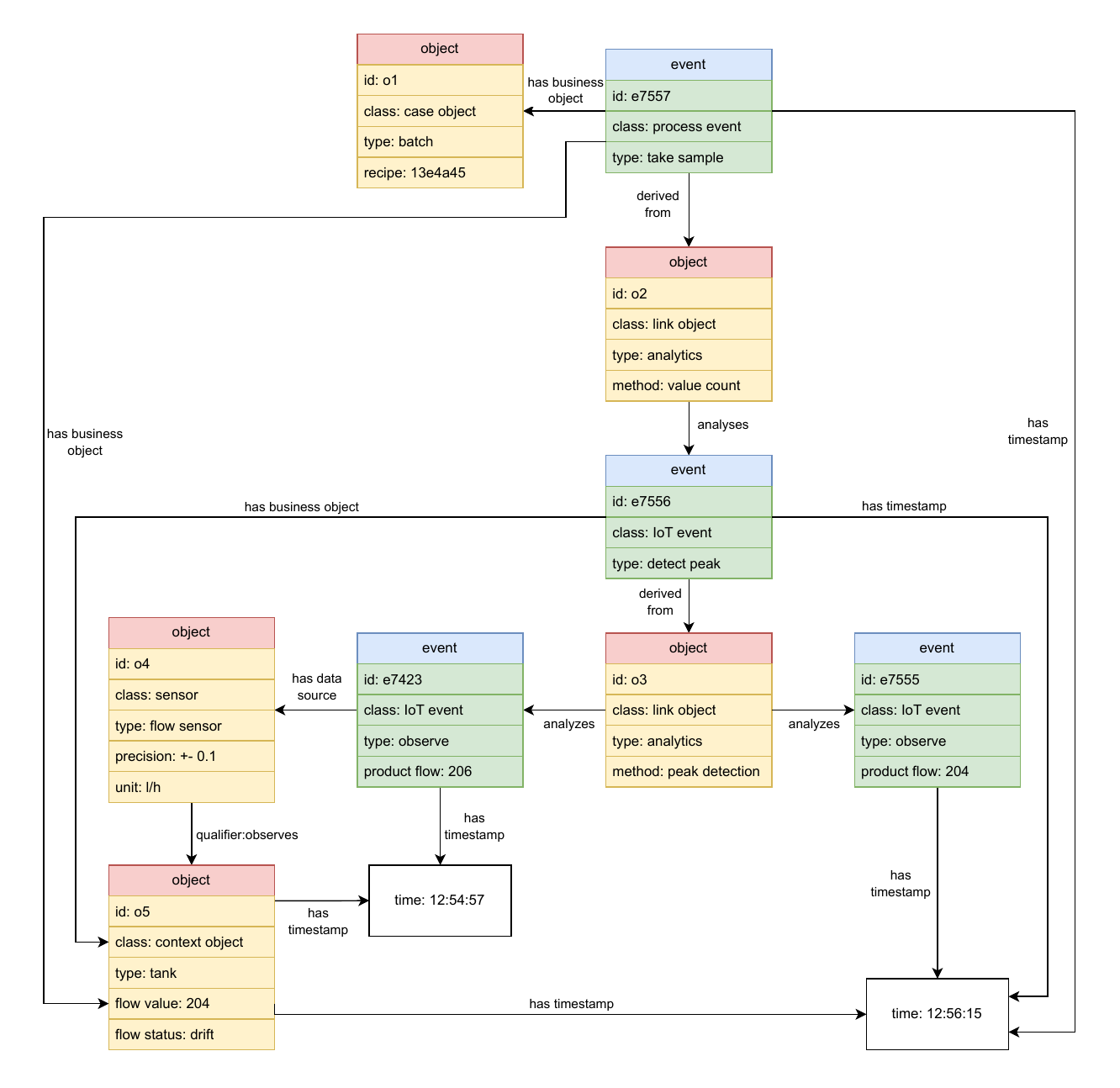}
    \caption{Instantiation of the \name for the manufacturing example. For readability and conciseness, we represented the instantiation as a UML class diagram, with objects and events as classes and attributes and attribute values as attributes (new).}
    \label{fig:manuexamp}
\end{figure}

To illustrate the use of our conceptual model, this section describes how the running example introduced in Section \ref{sec:example} can be represented in the \name. Figure \ref{fig:manuexamp} shows the instantiation of the model for the running example, while Table \ref{tab:exampleLog} shows a fragment of the events table of a CORE model log for the running example.


A Flow sensor object of type Sensor observes a Tank object, of type Context object, and makes observations of the product flow in the tank as IoT events, with the product flow value stored as an event attribute. Note that both the IoT event and the Tank object are related to the timestamp when a measurement happened to represent the change of object attribute value detected by the IoT event, in line with the OCEL 2.0 metamodel.

This shows an example of sensor data storage. By analyzing multiple consecutive sensor data observations, it is possible to derive higher level events. For example, three consecutive peaks in the product flow indicate that a sample of the product was drawn for analysis. This is represented as follows.

An Analytics object, of type Link, makes the junction between lower-level IoT events (observations) and a higher-level IoT event 'Peak detected', based on a peak detection algorithm. This event is also related to the Tank object. Then, a second Analytics Link object is used to represent the derivation of a process event 'Sample taken' based on the rule that three consecutive 'Peak detected' IoT events need to be recorded within a short time window (only one is shown in Figure \ref{fig:manuexamp} for simplicity). This process event has an activity label attribute and is related to a Batch Case object. Note that in the figure, derived higher-level events take the timestamp of the last lower-level event they are derived from.

\begin{table}[hbtp]
    \centering
    \caption{Example event log fragment including the data in Figure \ref{fig:manuexamp}.}
    \label{tab:exampleLog}
    \begin{tabular}{cc|cccc}
        Event ID & Object ID & Timestamp & Activity label & Event class & Value \\
        \hline \hline
        e1 & o1, o5 & 12:34:56 & Filtering & Process & Start \\
        \dots & \dots & \dots & \dots & \dots & \dots \\
        e7423 & o4, o5 & 12:54:57 & Observation & IoT & 206 \\
        e7424 & o4, o5 & 12:54:58 & Observation & IoT & 206 \\
        e7425 & o4, o5 & 12:54:59 & Observation & IoT & 207 \\
        \dots & \dots & \dots & \dots & \dots & \dots \\
        e7555 & o4, o5 & 12:56:15 & Observation & IoT & 204 \\
        e7556 & o2, o3 & 12:56:15 & Peak detected & IoT & Complete \\
        e7557 & o1, o2 & 12:56:15 & Take sample & Process & Complete \\
        e7558 & o4, o5 & 12:56:16 & Observation & IoT & 204 \\
        e7559 & o4, o5 & 12:56:17 & Observation & IoT & 202 \\
        \dots & \dots & \dots & \dots & \dots & \dots \\
    \end{tabular}
\end{table}

%% file: 6-implementation.tex
\section{Proof-Of-Concept Implementation}
\label{sec:implementation}
This section outlines the Python implementation of \name, outlining the architectural decisions and software components developed. It serves as a proof-of-concept\footnote{\hyperlink{https://github.com/chimenkamp/core-model-for-event-logs}{https://github.com/chimenkamp/core-model-for-event-logs}} and a practical contribution, enabling researchers and practitioners to apply \name to their specific use cases.
First, we provide an overview of the architecture and motivate our design decisions (\Cref{subsec:architecture_overview}). Furthermore, we describe the input layer, where the interfaces and parsers are defined (\Cref{subsec:input_layer}). Next, we report on the main logic, contained in the data transformation layer (\Cref{subsec:data_transformation layer}). Finally, we illustrate how the data can be represented using the OCEL standard, ensuring that our approach remains independent of any particular OCEL implementation (\Cref{subsec:ocel_representation_layer}).

\subsection{Architectural Overview}
\label{subsec:architecture_overview}
\begin{figure}[H]
    \centering
    \includegraphics[width=1\linewidth]{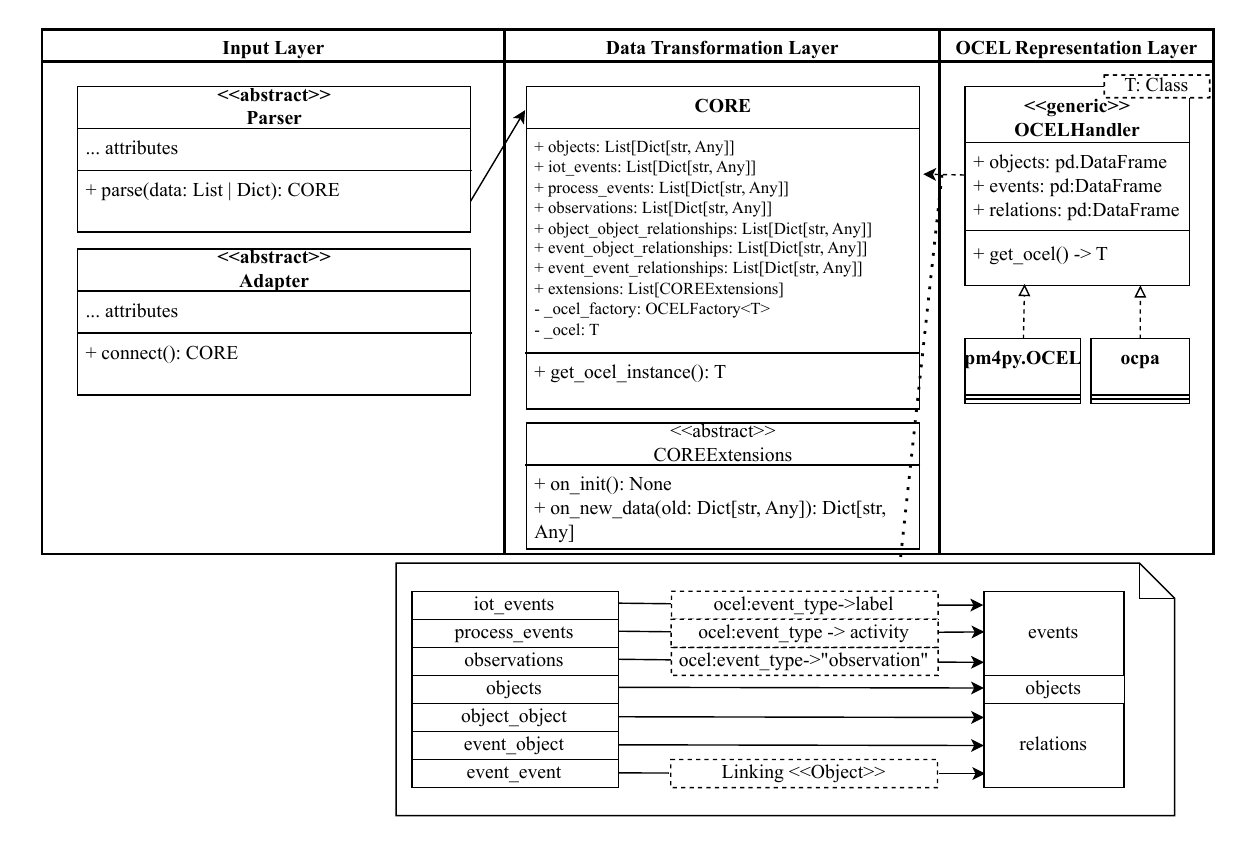}
    \caption{Overview of Layered Software Architecture}
    \label{fig:architecture}
\end{figure}

The decision of an architecture for \name adheres to specific requirements (e.g., model conformity, rapid prototyping). Model conformity ensures correctness and consistency. Model-driven engineering (MDE) allows quick iterations due to frequent metamodel changes. Given the dynamic \name development, we prioritize simplicity, reliability, and testability as key requirements for the architecture decision.
Based on these requirements and following established best practices~\citep{ford2020fundamentals}, we decided to use a \textit{Layered Software Architecture} as an architectural style for our implementation. Following this, we define three layers: the \textit{Input Layer} is necessary to parse the different existing logs (see \Cref{sec:relworks}). The \textit{Data Transformation Layer} links concepts between the input model and \name (see \Cref{subsec:link_to_other_models} and \Cref{fig:ocelModel}). Finally, the \textit{OCEL Representation Layer} transforms and represents our model in OCEL (see \Cref{fig:highLevelMetamodel}).
The architecture is visualized in \Cref{fig:architecture}, consisting of the three layers, Input Layer, Data Transformation Layer, OCEL Representation Layer. To closely reflect our implementation, we explicitly include Python-specific type annotations in the UML representations.

\subsection{Input Layer}
\label{subsec:input_layer}

\begin{figure}[H]
    \centering
    \includegraphics[width=0.5\linewidth]{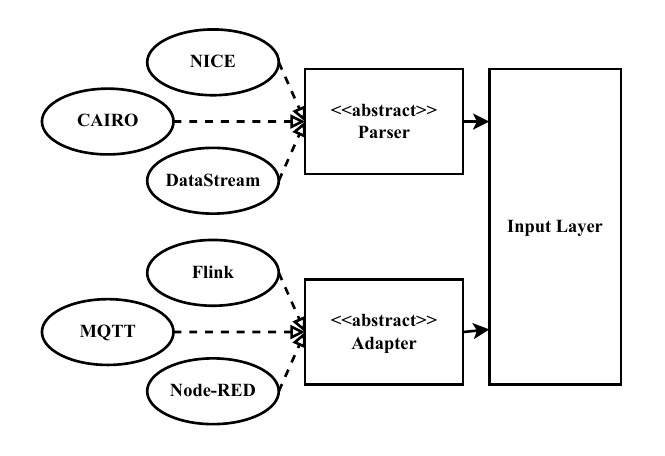}
    \caption{Input Layer}
    \label{fig:input_layer}
\end{figure}

The \textit{Input Layer} addresses the need for compatibility with the diverse set of existing models that \name must support (see \Cref{sec:relworks}). This compatibility ensures that the transformation process begins with a correct and semantically aligned interpretation of heterogeneous event formats. It consists of two main components (\Cref{fig:input_layer}). First, the abstract parser, handles the translation of existing IoT-enhanced event log formats (e.g., NICE, CAIRO, DataStream) with pre-defined classes. For that, the user will need to map the the components (i.e., from the custom format) to the \name components. In addition, it can be utilized to write custom parsers. This component defines a common interface for parsing different log formats into the internal data structures. To illustrate the concept transfer, we consider the DataStream parser as an example in the following. The DataStream format utilizes two different types of events: process events that represent high-level activities (e.g., “Start Production”, “Quality Check”), and stream points that capture IoT sensor observations. For each process event, the parser creates corresponding objects (e.g., machines, resources) and establishes relationships between them. For instance, when a machine operator starts a production process, this creates a process event with an associated resource object (the operator). Simultaneously, stream points are translated into observation objects and linked to the corresponding process events. Any values not directly required by \name (e.g., ‘system type’, ‘interaction type’) are preserved as attributes.

The second component is the abstract adapter, which connects real-world environments.   
This includes middleware platforms such as Flink, MQTT, and Node-RED. Therefore, adapters can be used to transform the incoming real-time data into the same internal format used by the parsers and consequently to the \name format. The continuous and unbounded nature of data streams poses unique challenges. Due to the potential volume of data, it is not possible to simply store the data in memory. Hence, we introduce \textbf{COREExtensions}, that can be used to adapt the behavior of \name. For example, the StreamingExtension introduces a caching mechanism that periodically writes accumulated data to binary files on disk. Furthermore, the extension maintains a lookup table to track long-term dependencies between events, objects, or events and objects when the original data has been cached. As an example, consider the following example: When a machine sends continuous sensor readings through Flink, the adapter creates observation events and temporarily holds them in memory. Once a certain threshold is reached (based on time or count), these observations are serialized to disk, but their relationships with process events and objects remain accessible through the lookup table.


\subsection{Data Transformation Layer}
\label{subsec:data_transformation layer}

\begin{figure}[H]
    \centering
    \includegraphics[width=0.75\linewidth]{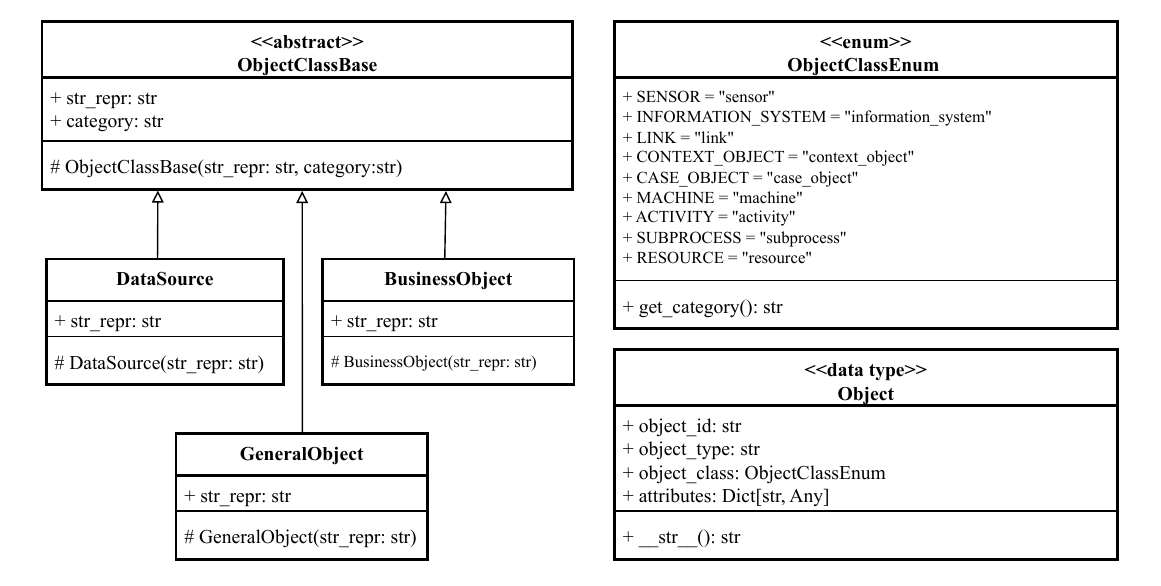}
    \caption{UML representation of the CORE object definition}
    \label{fig:object_implementation_uml}
\end{figure}

The \textit{Data Transformation Layer} focuses on transforming the CORE data structures into the OCEL 2.0 standard. This transformation step is critical, as it operationalizes the conceptual design (\ref{sec:model}) by aligning the internal semantics of the CORE objects, events, and relationships with the structural and syntactic constraints of OCEL 2.0. 
At its core, this layer transforms the \name classes (\Cref{fig:highLevelMetamodel}) into OCEL classes (\Cref{fig:ocelModel}). This transformation is performed in three phases:
The first phase transforms physical items, digital artifacts, and data sources from the \name into OCEL-compatible objects. As illustrated in \Cref{fig:object_implementation_uml}, the CORE implementation provides multiple utility classes. As described in \Cref{subsec:objects_taxonomy}, Core Objects are grouped into three categories (i.e., DataSources, BusinessObject, GeneralObject). This subdivision can be refined using the object\_class attribute of the object class. Moreover, the enum contains distinct possible object classes that are semantically connected to categorize. For example, a sensor is likely to be a data source. In addition to the object\_class attributes, the object\_type and object\_id attributes are assigned directly to the corresponding OCEL counterparts. Additionally, all use case-specific attributes are assigned in the columns ‘ocel:attr:[key]’, where the key represents the corresponding attribute label. Finally, each object\_id must be unique; otherwise, the object will not be transferred.

\begin{figure}[H]
    \centering
    \includegraphics[width=0.4\linewidth]{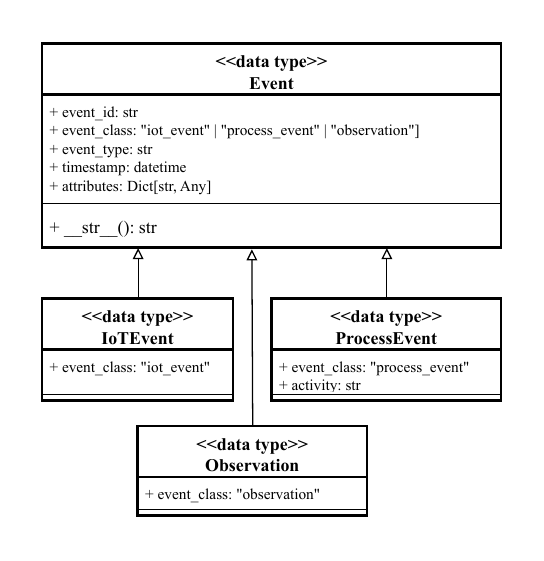}
    \caption{UML representation of the CORE event definition}
    \label{fig:event_implementation_uml}
\end{figure}

The second phase handles the event transformation. For that, the implementation provides an abstract Event class with three implementations (see \Cref{subsec:event_type_definition}, \Cref{fig:event_implementation_uml}). The event class requires four attributes, while the \textbf{ProcessEvent} implementation requires one additional activity attribute. The event\_id, timestamp, event\_class and attributes are mapped accordingly to the object implementation. The event\_type is dependent on the event\_class. First, if the event\_class is \emph{process\_event} the event\_type is the activity label. Next, if the event\_class is \emph{iot\_event} the event\_type is some meaningful label. Finally, if the event\_class is \emph{observation}, the event\_type is hardcoded to \emph{observed}. 

\begin{figure}[H]
    \centering
    \includegraphics[width=0.75\linewidth]{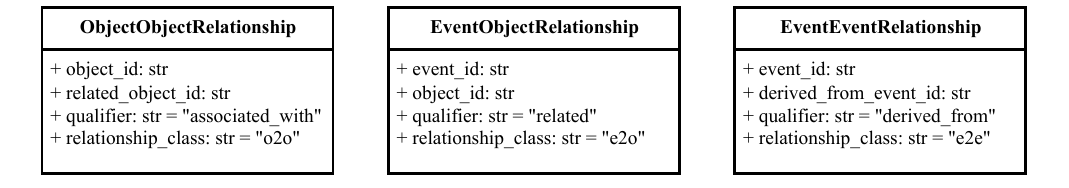}
    \caption{UML representation of the relationships}
    \label{fig:relationships_implementation_uml}
\end{figure}

The last phase establishes and maintains three relationships between the entities. As previously outlined, utility classes are also present in this phase (\Cref{fig:relationships_implementation_uml}). They define the required attributes for the different relationships. Generally, the (event/object)\_id is utilized for the mapping between the different entities. Each relationship is associated with a relationship\_class, which is also serialized in the relationship table. As there is a direct equivalent in the OCEL 2.0 standard for the e2o and o2o relationships, these can be transferred directly. The only exception is the mandatory activity that must be used for each relationship (in OCEL). As this is not required for every event\_class, the event\_type is used in those cases.
On the other hand, the event-to-event relationships require more parsing because there is not a corresponding definition in the OCEL standard. Given that, a linking object without semantically relevance is utilized. Therefore, two relationships are established. One between the first event and the linking object, and another between the linking object and the second event. Hence, the data can be queried as there are directly connected.

\subsection{OCEL Representation Layer}
\label{subsec:ocel_representation_layer}
The implementation should be independent of a specific OCEL implementation (e.g., pm4py.OCEL [\citep{pm4py}], ocpa [\citep{ocpa}]) and follow the structure provided in \Cref{fig:ocelModel}. This abstraction not only enhances the maintainability and flexibility of the implementation, but also ensures that the output remains reusable and interoperable.

To archive this, we utilize a bridge pattern, which separates the implementation (OCEL operations) from its abstraction (specific OCEL libraries). Therefore, the \textit{Data Transformation Layer} communicates with the OCELHandler class. It defines the high-level interface for OCEL operations, while different concrete implementations provide the actual functionality.

This modularity has significant implications for the future evolution of the \name. It ensures that the system can adapt to future developments or enhancements in OCEL implementations. It protects the implementation from becoming obsolete due to changes in external dependencies. Moreover, by facilitating easy switching between different OCEL frameworks, our approach significantly reduces the integration overhead typically associated with adopting new tools or libraries. Consequently, this architectural choice promotes long-term maintainability, supports a broader range of use cases, and fosters wider adoption within the community.

%% file: 7-evaluation.tex
\section{Evaluation}\label{sec:evaluation}
In this section, we evaluate the proposed \name by applying the metamodel to the use cases at the basis of the models introduced in Section~\ref{sec:background}. The evaluation serves a dual purpose, akin to demonstrating the soundness and completeness of the metamodel. First, we aim to demonstrate that the \name fulfills all requirements discussed in Sect.~\ref{subsec:define_objectives_of_a_solution}. Second, we aim to show that translating the original metamodel of each use case into the \name results in minimal information loss. We begin by discussing each use case and its metamodel translation individually. Then, since the use cases share common requirements, we assess the fulfillment of the requirements by considering the use cases altogether.

\subsection{Smart spaces NICE log}
\subsubsection{Use case description}
In this use case, we describe how we transcribe a log originally stored in the NICE format to our common \name. This synthetic event log, first presented in \citep{bertrand2026nice}, represents the behavior of two inhabitants living in a smart home for four weeks. To generate it, we used the smart spaces simulator presented in \citep{veneruso2023model}, which we adapted to generate NICE logs. More specifically, the two users were programmed to execute daily life activities such as \textit{Cook and eat}, \textit{Do the dishes}, \textit{Drink}, \textit{Eat cold}, \textit{Eat warm}, \textit{Exercise}, \textit{Go work}, \textit{Sleep}, \textit{Rest}, \textit{Shop}, \textit{Use Computer}, \textit{Watch tv} and \textit{Wc}. Moreover, the simulator was programmed so that during one week, the users show a different behavior, i.e., they stay at home the whole day instead of going to work in the morning and coming back home in the evening during weekdays. This change in behavior is due to extremely high outside temperatures during that week, which are tracked by temperature sensors and logged as IoT events.

\paragraph{Challenges of this use case}
\begin{itemize}
    \item Flexibility of the process: human routines are characterized by their context dependency, their lack of structure, the repetition of activities and the execution of multiple activities in parallel \citep{Leotta2015ApplyingPM};
    \item Structure of the NICE log: NICE logs are by design quite complex, with many pointers to other events and a nested structure.
\end{itemize}

\subsubsection{Translation to core log format}
The presence of many common concepts (and grounding in SOSA) make the translation of most events and objects relatively straightforward: IoT events become IoT events, etc.
However, some concepts did not have an equivalent in the \name and had to be translated. Data sources, for instance, became objects of class data source.
In addition to this, the structure is somewhat modified (e.g., no more data sources list, less strict separation between digital and physical worlds as in NICE, where most concepts have a digital and a physical variants, e.g., Objects can be Features of interest or Digital objects). More specifically:
\begin{itemize}
    \item Data sources become objects (of type data source)
    \item Bottom-level IoT events become Observations
    \item Context events disappear 
\end{itemize}

The translation of the NICE log into the \name format requires some considerations. In particular, the interrelated nature of the NICE format makes the translation unreasonable in terms of processing time. Therefore, the different elements (e.g., Features of interest, Sensors, or Context events) are translated concurrent. The Features of interests are translated into objects. The \name object type is derived from the NICE object type (location $\Rightarrow$ Business Object, date $\Rightarrow$ Business Object and user $\Rightarrow$ Resource). Next, the NICE Sensors are translated to objects with the object type Sensor and the attributes location and metadata. If there is a location given, an Object-Object-Relation is established with the aforementioned business object. As both metamodels have a concept of IoT event, Event-Object-Relationship and Observations, these can be directly translated.

\subsection{Manufacturing Use Case XES DataStream Logs}

\subsubsection{Use Case Description}

In manufacturing processes, multiple (IoT) devices operate collaboratively to produce a part. These processes encompass various steps, including the handling of parts, which may involve multiple machines, transportation between different machines, and the measuring of parts.
In the following, two use cases are presented that demonstrate the integration of IoT devices into manufacturing processes. In each case, the processes are managed by a process execution engine and are logged using the XES DataStream extension, ensuring real-time tracking and analysis of production events. This approach facilitates enhanced process monitoring, optimization, and traceability, contributing to improved efficiency and product quality.~\citep{datastream}:
\begin{itemize}
    \item In the IoT lab at Trier University\footnote{More information can be found at \url{https://iot.uni-trier.de}.}, a Fischertechnik learning factory is used to carry out research related to BPM \citep{Malburg.2020_BPMAndIoT}. The use case includes machines which are equipped with several types of sensors that collect production-related data, such as the state of connected light barriers or switches. Complementing this Internet of Things (IoT) data, information pertaining to process executions is systematically recorded and maintained as event logs, facilitating comprehensive monitoring and analysis of manufacturing processes \citep{Grueger.2023}. 
    \item Another use case\footnote{\url{https://doi.org/10.5281/zenodo.7958478}} is a process where a machine tool, a robot, and a measuring machine work together to produce a rook, transport it to the measuring machine, and finally put it on a tray. Multiple parts are produced in this way. Furthermore, data about performed actions as well as additional data available during the process (e.g., data measured by sensors during the machining of a part, diameter measurements at different points on the part, data about the movement of the robot, ...) is logged.
\end{itemize}
Both of these use cases share similar features in the areas of (i) integrating (IoT) devices into processes that are later executed using a process engine, (ii) collecting data from relevant sensors for the use case in addition to actively controlling the machines, and (iii) combining information that is different regarding its granularity (i.e., high-level process event vs. low-level IoT event) and in frequency of occurrence, ranging from infrequent events recorded once per minute to high-frequency data captured at rates of several thousand events per second, into a unified event log.

\paragraph{Challenges}
Based on the features discussed above, the following challenges arise for the described use cases:
\begin{itemize}
  \item Sensor data must be integrated into the process log taking into consideration (i) the high frequency of sensor data, (ii) the immediate availability of sensor data (independent of long-running activities) to enable analysis as close as possible to the measurements, and (iii) the connection of sensor data to the activities in which they have been recorded.
  \item When enacting a process, the process model is already known which means that additional information about the collection of sensor data or how data in general will be used throughout the remaining process (i.e., decision-making, as input for subsequent activities, ...). This should also be represented in the log.
  \item Due to the process controlling all actions, sensor data collection must be triggered by the process itself, which means it has to be contained in the process model explicitly. This ensures synchronization between process execution and the event log.
\end{itemize}

\subsubsection{Translation to core log format}
The logs for the use cases described before are logged using the XES DataStream extension~\citep{datastream}. In the following, it is described how they can be translated into the core model format:
\begin{itemize}
  \item The translation between the Trier University use case and the \name has been conducted as follows: First, to model process participants the DataStream resource (i.e., "org:resource") is parsed into an object with the type and class set to resource. Thereafter, the “stream:points” are analyzed. If the current “stream:point” contains an "concept:name" attribute, a Process Event is derived. The remaining "stream:points" are translated into Observations including the following attribute: "system", "system\_type", "observation", "procedure\_type" and "interaction\_type" and "value". Lastly, Process Events are linked to their executing resource using an Event-Object-Relationship and consequently, the process events are linked to the IoT-derived observations. 
  \item The transition from the TUM use case (i.e., DataStream) to the \name can be summarized as follows: IoT-Devices from DataStream (i.e., "stream:name", “stream:source”) are translated into objects with the class "machine" and the type IoT-Device. “stream:point” are events of class observation and are linked to the corresponding IoT-Device with an Event-Object-Relationship. If the DataStream event contains a "concept:name" it is parsed into a Process Event. Lastly, an Event-Event-Relationship between the Process Event and the corresponding IoT-Events is established.
\end{itemize}

\subsection{Blood donation process use case (CAIRO)}
\subsubsection{Use case description}
In this use case, the focus is on the IoT-supported online monitoring and conformance checking of the blood donation process, augmented with hand hygiene guidelines from the World Health Organization (WHO) that prescribe additional process steps of hand disinfection~\citep{franceschetti2023proambition}. The goal is to be able to provide a timely feedback to the healthcare workers regarding the conformance with the hand hygiene requirements, reducing the risk of contamination that could threaten the patients' health. As the process is fully manual and executed without the support of a process engine, the challenge here is to be able to generate in real-time IoT events from the process execution and abstract them into the respective process events. However, ambiguities deriving from multiple process events such as \textit{Apply tourniquet} and \textit{Insert needle} being represented by the same IoT events pose a challenge to the event abstraction. This results in multiple process events being generated from a given set of IoT events, with attached a quantification of the respective ambiguity.

\paragraph{Challenges of this use case}
\begin{itemize}
    \item Identifying and representing ambiguity: in the IoT event to process event abstraction, ambiguity is introduced due to the lack of process-awareness of the IoT events. This makes it challenging to abstract IoT events into the correct process events.
    \item Online conformance checking with ambiguity: due to the ambiguity introduced in the IoT event to process event abstraction, there is uncertainty regarding which process events to consider for conformance checking against a known process model.
\end{itemize}


\subsubsection{Translation to core log format}
Several concepts are present in the \name. Traces can be mapped to objects with the case class. Each event (i.e., “stream:point”) is translated into an IoT-Event with its corresponding attributes. Subsequently, an Event-Object-Relationship between the case object and all associated IoT-Event is established. 
Given that, ambiguity in this relation is specific to this use case and cannot be explicitly represented in the \name and requires an extension of it.


\subsection{Requirements Fulfillment in the Use Cases}\label{sect:requirements_fulfillment}
In this section, we take a closer look at how the requirements listed in Section \ref{subsec:define_objectives_of_a_solution} are addressed by our model at the hand of the use cases to validate the model.

\begin{itemize}
  \item R1 (``Different Granularity Levels'') is a requirement for all use cases, as they all handle IoT events and process events, which can both be stored in the \name (see description of translation provided above).
  \item R2 (``Context Independent from Control-flow'') is mainly relevant to the smart spaces use case, where context parameters detected by sensors should be updated independently from process events. The \name can do this by updating object attributes via IoT events.
  \item R3 (``Traceability'') is required by all use cases. This is addressed by the \name via link events, which connect IoT events to the process events responsible for their collection in the manufacturing use case, and relate process events to the IoT events they are abstracted from in the smart spaces and healthcare use cases.
  \item R4 (``Semantic Annotations'') is a requirement of the manufacturing and of the healthcare use cases. In the manufacturing use cases, semantic annotations are present in both logs. In the first log, semantic information from ontologies is used to describe behavior more precisely and to enhance its semantic richness. In the second manufacturing use case metadata such as local timestamps from the machine or information about the source sensor data are included in the original logs. Finally, in the healthcare use case, a known normative process model is stored as a semantic annotation.
  \item R5 (``Flexible Case Notion'') is a requirement of the smart spaces use case, which is fulfilled by the \name with business objects, which can represent flexible case notions.
  \item R6 (``Different Data Sources'') are needed by the smart spaces and manufacturing use cases as data are provided by different machines, sensors and information systems. This requirement is addressed with the information system and sensors data sources.
  \item R7 (``Represent Metadata'') In the second manufacturing use case, semantic metadata is utilized to log updates to the model. The \name can store these annotations using appropriate objects to connect the events to.
\end{itemize}

%% file: 8-discussion.tex
\section{Discussion}
\label{sec:discussion}
The set of requirements presented in Sect.~\ref{sec:method} is derived from the harmonization of requirements from different models addressing diverse use cases. As such, it provides a solid foundation for the development of a suitable representation of IoT-enhanced process logs. As reported in Sect.~\ref{sec:evaluation}, the proposed \name successfully fulfills these requirements, effectively supporting different use cases in various IoT domains. Moreover, the application of the proof-of-concept implementation to \textit{concrete} use cases underscores the suitability of the metamodel for multiple near real-world applications.

The \name offers the following implications for researchers:

\begin{itemize}
\item \textbf{Unified Representation:} The model provides a standardized approach to integrating IoT data, process data, and contextual information across different levels of granularity (cf. Sect.~\ref{sec:evaluation}). Thus, the \name embodies a common reference point for future research, facilitating cross-study comparisons, meta-analyses, and the development of new process mining algorithms.
\item \textbf{Flexibility and Extensibility:} By supporting semantic annotations, metadata representation, and diverse data sources, the model accommodates a wide range of use cases that require the representation of additional data beyond events, as demonstrated in the evaluation in Sect.~\ref{sec:evaluation}. This extensible architecture paves the way for future extensions and customized adaptations, for example capturing domain-specific knowledge. We expect this to be instrumental to enable further advancements in context-aware as well as semantics-aware process mining~\citep{rebmann2022enabling}.
\item \textbf{Interoperability:} We deliberately made the choice of implementing the \name within the OCEL 2.0 format~\citep{berti2024ocel}. Currently, OCEL 2.0 is the go-to format adopted by researchers in process mining~\citep{goossens2024object}. Therefore, we expect that this will facilitate uptake by researchers and integration with existing use cases, fostering data exchange and reuse.
\item \textbf{Interdisciplinary Research Opportunities:} The \name is positioned at the intersection of IoT and process mining. This encourages interdisciplinary research efforts, where experts in each area can collaboratively address challenges such as data fusion, real-time monitoring, and predictive analytics in complex process environments.
\end{itemize}

Moreover, the \name yields the following implications for industry practitioners:

\begin{itemize}
    \item \textbf{Data Exchange and Integration:} The adoption of the \name allows organizations to translate and integrate data from various sensors, information systems, and legacy event log formats. The definition of mappings between their existing IoT and process-related data and the \name facilitates data exchange and integration between different systems.
    \item \textbf{Enhanced Process Analytics:} The metamodel design allows for the simultaneous representation of process events and IoT observations. This duality enables practitioners to bridge the gap between the physical world and digital process representations. This enhances the ability to monitor and analyze business processes at different levels of granularity, thereby obtaining more sophisticated insights into enterprise operations.
    \item \textbf{Reduction of Integration and Adaptation Costs:} A standardized metamodel reduces the need for custom data cleaning and integration efforts. Organizations can more easily adopt advanced process mining tools and analytics platforms that are built around such common standards, lowering both operational and development costs over time.
\end{itemize}

Overall, the \name addresses several challenges, providing benefits for researchers and practitioners. Still, we recognize some limitations of the metamodel. Specifically, the reconciliation of diverse requirements from different use cases and research groups resulted in a number of trade-offs in the expressive power of the metamodel. This results in the lack of \textit{explicit} concepts in the \name able to represent use case-specific concepts, such as the Formal Model concept from CAIRO, which represents a formal process specification and is relevant in the smart healthcare use case. Nevertheless, the \name remains open for specialization and extension to ensure that also use-case specific requirements can be met. This can be achieved, for example, by leveraging event attributes or by specializing the general object concept and its associations (cf. Sect.~\ref{sec:model}). This extensibility can be supported by a model-driven engineering (MDE)~\citep{kent2002model} approach, where the \name acts as a central specification from which new artifacts can be derived. With this approach, adopting the metamodel and introducing new subclasses or domain-specific concepts does not require a full re-implementation, as changes can be integrated (semi-)automatically using MDE tools. We emphasize that these trade-offs reflect deliberate design choices made during the metamodel development to prioritize generalizability and interoperability--which we consider major goals of the metamodel.

We also acknowledge threats to the validity of the present contribution. First, we recognize the presence of potential bias in the gathering of requirements. The design of the metamodel by researchers purely from academia may have introduced bias due to the focus on specific use cases and research goals. This may lead to potentially excluding some niche use cases or industry-specific requirements. However, this threat is mitigated by the relatively high number of researchers involved and their different backgrounds and collaborations with industry partners. Additionally, the number and diversity of use cases considered for the metamodel evaluation also mitigates this threat. The evaluation scope constitutes a second threat to the validity, since the use cases considered may not fully capture all possible scenarios of IoT-BPM integration. However, the diversity of the use cases mitigates this threat by covering a range of domains exhibiting different requirements and characteristics.

%% file: 9-conclusion.tex
\section{Conclusion and Future Work}
\label{sec:conclusion}
In conclusion, the contribution of this paper is manifold: 1) we present a set of requirements, important for the development of a suitable representation format for IoT-enhanced event logs, 2) we propose a new model referred to as \name for IoT-enhanced event log, rooted in the object-centric paradigm to foster data exchange in the community, 3) we present our prototypical Python implementation of this new model, compatible with the PM4PY package and the OCEL 2.0 specification, and 4) we use this implementation for evaluation in several near real-world IoT use cases. To ensure the quality of the model and rigor in its design, the model has been created following the design science research methodology, building on existing IoT ontologies and IoT-enhanced log models, and based on a set of requirements gathered from experts in IoT-enhanced PM. Furthermore, the application of the metamodel to multiple use cases indicates that existing logs in various IoT-enhanced log formats can easily be translated into logs in our \name format, demonstrating the achievement of the interoperability goal of the model.

In future work, we plan to improve the usability of our new model. To do so, we will first aim to offer our \name as an official extension of the OCEL 2.0. Then, we would like to integrate our Python implementation in a future release of the PM4PY package to make it widely available to researchers and end users. This way, we will also work further towards the objective of interoperability of our model.